\documentclass[a4paper,12pt]{scrartcl}
\usepackage[ngerman, english]{babel}
\usepackage[T1]{fontenc}
\usepackage{amsmath}
\usepackage{arydshln}
\usepackage{multirow}
\usepackage{caption}
\usepackage[authoryear]{natbib}
\usepackage{amssymb}
\usepackage{verbatim}
\usepackage{amsthm}
\usepackage[section]{placeins}
\usepackage{paralist}
\usepackage{graphicx}
\usepackage{subfigure}

\DeclareMathOperator{\E}{E}

\DeclareMathOperator{\AIC}{AIC}
\DeclareMathOperator{\BIC}{BIC}

\DeclareMathOperator{\cM}{\mathcal{M}}
\DeclareMathOperator{\trace}{tr}

\DeclareMathOperator{\td}{td}
\DeclareMathOperator{\mse}{MSE}
\DeclareMathOperator{\amse}{AMSE}
\DeclareMathOperator{\mmse}{MMSE}

\numberwithin{table}{section}
\numberwithin{equation}{section}
\numberwithin{figure}{section}

\begin{document}
\title{Model Selection versus Model Averaging in Dose Finding Studies}
\author{
 {\small  Schorning, Kirsten} \\
{\small Ruhr-Universit\"at Bochum} \\
{\small Fakult\"at f\"ur Mathematik} \\
{\small 44780 Bochum, Germany} \\
{\small e-mail: kirsten.schorning@rub.de}\\
\and
\small{Bornkamp, Björn}\\
{\small Novartis Pharma AG}\\
{\small Lichtstrasse 35} \\
{\small 4002 Basel, Switzerland}\\
{\small e-mail:  bjoern.bornkamp@novartis.com}\\
\and
{\small Bretz, Frank}\\
{\small Novartis Pharma AG}\\
{\small Lichtstrasse 35} \\
{\small 4002 Basel, Switzerland}\\
{\small e-mail:  frank.bretz@novartis.com}\\
\and
{\small Dette, Holger}\\
{\small Ruhr-Universit\"at Bochum} \\
{\small Fakult\"at f\"ur Mathematik} \\
{\small 44780 Bochum, Germany} \\
{\small e-mail: holger.dette@rub.de}
}
\date{}

\maketitle

\begin{abstract}
Phase II dose finding studies in clinical drug development are typically conducted to adequately
  characterize the dose response relationship of a new drug. An important decision is then on the choice of a suitable dose response
  function to support dose selection for the subsequent Phase III studies. In this paper we compare different approaches
  for model selection and model averaging using mathematical properties as well as simulations. Accordingly, we review and
  illustrate asymptotic properties of model selection criteria and
  investigate their behavior when changing the sample size but keeping
  the effect size constant. In a large scale simulation
  study we investigate how the various
  approaches perform in realistically chosen settings. Finally, the different methods are illustrated with a
  recently conducted Phase II dose-finding study in patients with chronic obstructive pulmonary disease.
\end{abstract}

\bigskip
\noindent
Keywords and Phrases: Model selection; model averaging; clinical trials; simulation study

\section{Introduction}

A critical decision in pharmaceutical drug development is the
selection of an appropriate dose for confirmatory Phase III clinical
trials and potential marketing authorization.  For this purpose, dose
finding studies are conducted in Phase II to investigate the dose
response relationship of usually $3-7$ active doses in the intended
patient population for a clinically relevant endpoint; see
\cite{ting:2006} among many others.

Traditionally, dose response studies were analyzed by treating dose as
a categorical variable in an analysis-of-variance (ANOVA) model. Only
in the past 20 years the use of regression modeling approaches where
dose is treated as a quantitative variable has become more popular. We
refer to, for example, \cite{bret:hsu:pinh:2008} for an overview of
both approaches, and the White Paper of the Pharmaceutical Research
and Manufacturers of America (PhRMA) working group on adaptive dose
ranging studies (\cite{phrma:2007}) for a comparison of different ANOVA
and regression-based approaches.

If a non-linear regression model is adopted, a natural question is
which regression (i.e. dose response) function to utilize.  This
becomes even more important in the regulated context of pharmaceutical
drug development, where the employed regression model should be
pre-specified at the design stage. This specification thus takes place
at a time, when only limited information is available about the dose
response relationship, resulting in model uncertainty. Several authors
(e.g. \cite{thom:2006, drag:hsua:padm:2007}) argued that a flexible
monotonic model, such as an Sigmoid Emax model, can be used for all
practical purposes, as it approximates the commonly observed dose
response shapes well. While generally applicable, this flexible model
can sometimes be challenging to fit with a small number of doses. In
addition, while several models might fit the data similarly well, due
to the often sparse data they might still differ on certain estimated
quantities of interest, e.g. the target dose estimate.

The MCP-Mod method
(see \cite{bret:pinh:bran:2005,pinh:born:glim:2014,ema:2014}) tries to
address the model uncertainty problem by acknowledging it explicitly
as part of the methodology.  The main idea is to determine a candidate
set of dose response models at the trial design stage.
After completing the trial one
either selects a single dose response function out of the candidate
model set or applies model averaging based on the individual model
fits.  Thus, the MCP-Mod approach allows one to employ either model
selection or model averaging.  \cite{verr:2014} discussed by means of
two real examples their experiences on how to proceed with model
selection and model averaging using MCP-Mod in practice.

Model selection has the advantage that it results in a single model
fit, which eases the interpretation and communication. But it is also
known that selecting a single model and ignoring the uncertainty
resulting from the selection will result in confidence intervals with
coverage probability smaller than the nominal value, see for example
\cite{born2015} for a high-level discussion or Chapter 7 in
\cite{clae2008} for a mathematical treatment.  A partial solution to
this problem is to use model averaging. By acknowledging model
uncertainty explicitly as part of the inference one will typically
obtain more adequate (i.e usually wider) confidence intervals. There
exists empirical evidence that model averaging also improves the
estimation efficiency (see \cite{raft:2003} or \cite{brei1996}), even
though authors did not consider dose-finding setting in particular.

The purpose of this paper is to investigate and compare different
model selection and model averaging approaches in the context of Phase
II dose finding studies.  Accordingly, we introduce in
Section~\ref{sec:case} a motivating case study to illustrate the
various approaches investigated throughout this paper.  Next, we
briefly review the mathematical background of different selection
criteria and compare them with respect to some of their asymptotic properties
in Section~\ref{sec:not}.  In Section~\ref{sec:sim}, we describe the
results of an extensive simulation study.  We revisit the case study
in Section~\ref{sec:caserev} and provide some general conclusions in
Section~\ref{sec:concl}.

\section{A Case Study in Chronic Obstructive Pulmonary Disease (COPD)}\label{sec:case}

\begin{figure}[t!]
\centering
\includegraphics[width=0.55 \textwidth]{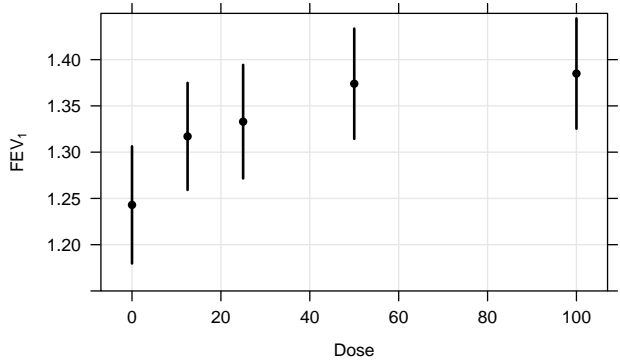}
\caption[Example]{\emph{Mean responses and marginal 95\% confidence intervals for the COPD case study.}}
\label{fig:data}
\end{figure}

This example refers to a Phase II clinical study of a new drug in
patients with chronic obstructive pulmonary disease (COPD). The
primary endpoint of the study was measured through the forced
expiratory volume in one second ($\mbox{FEV}_1$) measured in liter,
after 7 days of treatment.  The objective of this study was to
determine the dose response relationship and the target dose that
achieves an effect of $\delta$ over placebo. In COPD an improvement
$\delta$ of $0.1-0.14$ liters on top of the placebo response are
considered clinically relevant. To this end, four active dose levels
(12.5, 25, 50 and 100 mg) were compared with placebo. Point estimates
and standard errors for the treatment groups resulting from an ANCOVA
fit are available from clinicaltrials.gov (NCT00501852). The original
study design was a four-period incomplete block cross-over study; see
also \cite{verk2010}.  For the purpose of this article we simulated a
parallel group design of 60 patients per group (thus 300 patients in
total), so that the point estimates and standard errors match the
reported estimates exactly.
Figure~\ref{fig:data} displays the mean responses at the
five dose levels (including placebo) together with the marginal
95\% confidence intervals.

For our purposes, we assume that five candidate models had been
identified at the design stage to best describe the data after
completing the trial.  More specifically, we assume the five
dose-response functions summarized in Table~\ref{tab:candmod}, namely
the linear, quadratic, Emax, Sigmoid Emax and ANOVA model; see
Section~\ref{sec:not} for the notation used in
Table~\ref{tab:candmod}. The questions at hand are (i) which of these
candidate models should be used for the dose response modeling step,
(ii) whether model selection or averaging should be used, and (iii)
which specific information criteria should be employed to perform
either model selection or averaging. We will revisit and analyse this
case study in Section~\ref{sec:caserev}.

\begin{table}[t!]
\centering
\begin{tabular}{|c|c|c|r|}
\hline
Number & Model & Function $\eta(\cdot, \theta)$ & Parameter specifications\\
\hline
1 & Linear & $\vartheta_0 + \vartheta_1 d$  & $\theta_{1}=(0,-1.65/8)$ \\
2 & Quadratic & $\vartheta_0 + \vartheta_1 d +{\vartheta_2} d^2$ & $\theta_{2}=(0,-1.65/3, 1.65/36)$ \\
3 & Emax &$ \vartheta_0 + \frac{\vartheta_1 d}{\vartheta_2+d}$ & $\theta_{3}=(0,-1.81, 0.79)$ \\
4 & Sigmoid Emax &$\vartheta_0 + \frac{\vartheta_1 d^{\vartheta_{3}}}{\vartheta_2+d^{\vartheta_{3}}}$ & $\theta_{4}=(0,-1.7,4,5)$  \\
5 & ANOVA  & $\eta(d_i, \theta)=\vartheta_i, \, \, i=1, \ldots, k $ & $\theta_{5}=(0, -1.29, -1.35, -1.42,-1.5,$ \\
  &  &  & $ -1.6, -1.63, -1.65, -1.65)$\\
\hline
\end{tabular}
\caption[Different Dose Response Models]{\textit {The five candidate dose response models utilized in the case study, together with the parameter specifications used in the simulation study from Section~\ref{ssec:des}.}}
 \label{tab:candmod}
\end{table}

\section{Model Selection and Model Averaging}\label{sec:not}
We assume $k$ different dose levels $d_1, \ldots, d_k$, where often
$d_1=0$ is the placebo. The set $\Xi= (d_1, \ldots, d_k)$ of
$k=k(\Xi)$ dose levels is called design throughout this paper.  We
further assume that for each dose level $d_i$ we have $n_i$ patients
$i=1, \ldots, k$, where $N=\sum_{i=1}^k n_i$.  The individual
responses are denoted by
\begin{equation} \label{eq:sample}
y_{11}, \ldots, y_{1n_1}, \ldots, y_{k1}, \ldots, y_{k n_k}.
\end{equation}

Throughout this paper, we assume that the observations in \eqref{eq:sample} are realizations of random variables $Y_{ij}$ defined by
\begin{equation} \label{eq:model}
Y_{ij}= \eta(d_i, \theta) + \varepsilon_{ij} \quad  j=1, \ldots, n_i, \, i= 1, \ldots, k,
\end{equation}
where $\varepsilon_{11}, \ldots, \varepsilon_{kn_k}$ are independent
and normally distributed random variables, i.e. \linebreak
$\varepsilon_{ij} \sim~\mathcal{N}(0, \sigma^2)$. Here, $ \eta(d_i,
\theta)$ denotes the mean response at dose $d_i$. The competing dose
response mean functions in \eqref{eq:model} are denoted by $
\eta_\ell(d, \theta_\ell)$, $\ell=1, \ldots, L$.  For example, in the
case study presented in Section~\ref{sec:case} we assumed the $L=5$
candidate models $\mathcal{M}_1, \ldots, \mathcal{M}_5$, summarized in
Table \ref{tab:candmod}.

In the remainder of this section we give a brief overview of commonly used information criteria for selecting a model from  a given class of competing models.
All criteria can be represented in the form
\begin{equation} \label{eq:allcrit}
2 \max_{\theta_\ell} \log \mathcal{L}_N (\cM_\ell, \theta_\ell) - 2 \, pen_{\ell, I}
\end{equation}
where $\mathcal{L}_N$ denotes the likelihood function and $pen_{\ell,
  I}$ a penalty term which differs for the different models $\cM_\ell$
and selection criteria $I$.  Table \ref{tab:msc} summarizes the
penalty terms of different criteria that will be introduced below and
investigated in later sections.

\begin{table}[t!]
\centering
\begin{tabular}{cc}
\hline
Model Selection Criterion $I$ & Penalty Term for model $\cM_\ell$ \\
\hline
AIC & $ d_{\cM_\ell}$ \\
\hline
$\mbox{AIC}_C$ & $ \frac{N d_{\cM_\ell}}{N-d_{\cM_\ell}-1} $\\
\hline
BIC & $ 0.5 \log(N) d_{\cM_\ell} $\\
\hline
$\mbox{BIC}_2$ & $ 0.5 (\log(N) d_{\cM_\ell}  -  \log(2\pi) d_{\cM_\ell})$ \\
\hline
TIC & $\trace(\hat{J}^{-1} \hat{K})$ \\
\hline
\end{tabular}
\caption[Considered model selection criteria]{\textit{Five model selection criteria and their corresponding penalty terms investigated in this paper, where $d_{\cM_\ell}$ denotes the dimension of the parameter for model $\cM_\ell$.}}
\label{tab:msc}
\end{table}

\subsection{Information criteria based on the AIC}\label{sec:aic}
AIC-based information criteria are often motivated from an information
theoretic perspective.  Let $g(y|d)$ denote the true but unknown
density of the response variable $Y$ given the dose $d$.  In order to
estimate target doses of interest and the dose response curve, we want
to identify a model $\cM$ defined by a parametric density $p_{\cM} (y
| d, \theta_{\cM})$ for the response variable $Y$ from a given class
of $L$ parametric models which approximates the true density $g(y|d)$
best.  In order to measure the quality of the approximation we use the
Kullback Leibler divergence (KL-divergence)
\begin{equation}\label{eq:KL-dist}
KL(p_{\mathcal{M}}, g)= \sum_{i=1}^{k} \frac{n_i}{N} \int \log\left( \frac{g(y|d_i)}{p_{\mathcal{M}}(y| d_i, \theta_{\mathcal{M}})} \right) g(y|d_i) d y.
\end{equation}
The KL-divergence serves as a distance measure between densities.
It is nonnegative and equal to zero if $g(y|d)=
p_{\mathcal{M}}(y| d, \theta_{\mathcal{M}})$.
Based on the KL-divergence a model $\mathcal{M}$  from a given class of $L$ models, say $\cM_1, \ldots, \cM_L$, is called the best approximating model if its density (with corresponding optimizing parameter $\theta^*_{\mathcal{M}})$ minimizes the KL-divergence to the true density $g(y| d)$ compared to the KL-divergence of the other $L-1$  models.

In practice, the identification of the best approximating model within a set of $L$ candidate models  $\cM_1, \ldots, \cM_L$ by minimizing the KL-divergence \eqref{eq:KL-dist} is not possible because this criterion depends on the unknown true density.
However the divergence and the parameters corresponding to the best approximation can be estimated from the data $y_{11}, \ldots, y_{1n_1}, \ldots, y_{k1}, \ldots, y_{kn_k}$.
Ignoring the terms that are the same across all models, one hence needs to minimize
$$ Q_N(\cM_\ell):= \E \left[ \sum_{i=1}^k \frac{n_i}{N} \int  \log
  p_{\ell}(y| d_i, \hat{\theta}_{\ell} )g(y|d_i)dy \right], $$ where the
expectation is taken with respect to the distribution of the maximum likelihood
(ML) estimator $\hat{\theta}_{\ell}$ of
the parameter $\theta_\ell$ in model $\cM_\ell,\ell=1, \ldots, L$. \\
\noindent
It is known that the empirical estimator of this quantity, the log
likelihood $\frac{1}{N} \max_{\theta_\ell}
\log(\mathcal{L}_N(\cM_\ell, \theta_\ell))$ is a biased estimator and
overestimates $Q_N(\cM_\ell)$, leading to overfitting (cf. 
\cite{clae2008}).
A bias corrected estimator instead
is given by
\begin{equation}\label{eq:est}
Q^*_N(\cM_\ell)= \frac{1}{N}  \max_{\theta_\ell} \log(\mathcal{L}_N(\cM_\ell, \theta_\ell))  - \frac{pen_{\ell,I}}{N}.
\end{equation}
Using different estimators for the penalty term thus leads to different
model selection criteria; see Table \ref{tab:msc}.  \cite{clae2008}
discussed under which circumstances the different penalties lead to an
approximately unbiased estimation of $Q_N(\cM_\ell)$, in Appendix
\ref{sec:append-backgr-aic} we provide further technical background on
asymptotic approximations of the bias term.  Setting $pen_{\ell,I} =
d_{\cM_\ell} $ leads to the popular Akaike information criterion (AIC;
\cite{akai1973})
\begin{equation*} \label{AIC}
\AIC(\cM_\ell)=  2 \max_{\theta_\ell} \log(\mathcal{L}_N(\cM_\ell, \theta_\ell)) - 2 d_{\cM_\ell}.
\end{equation*}
The coefficient $2$ is added because of approximation arguments (see
among others \cite{clae2008}).  \cite{hurv1989} pointed out that the
dimension $d_{\cM_\ell}$ is not a good estimator of the bias for small
sample sizes and proposed the penalty term
$\frac{Nd_{\cM_\ell}}{N-d_{\cM_\ell}-1}$, leading to the corrected AIC
($\AIC_C$).  Also, \cite{take1976} suggested the penalty term
$\trace(\hat{J}^{-1}(\cM_\ell) \hat{K}(\cM_\ell) )$, leading to
Takeuchi's or the Trace Information Criterion (TIC), where $K$ denotes
the Fisher information matrix and $J$ the negative inverse of the
expectation of the second derivative of the log likelihood function.
Both $K$ and $J$ are estimated; see Appendix
\ref{sec:append-backgr-aic} for details.

\subsection{Information criteria based on the BIC}\label{sec:bic}
Roughly speaking, the Bayesian Information Criterion (BIC; \cite{schw1978}) chooses the most likely model based on the data. More precisely, let $Pr(\cM_1), \ldots, Pr(\cM_L)$ denote the prior probabilities for the models $\cM_1, \ldots, \cM_L$ and $p_{1}(\theta_1), \ldots, p_{L}(\theta_L)$ prior distributions for the corresponding parameters $\theta_1, \ldots, \theta_L$, respectively. Using Bayes' theorem and the observations $y^N= (y_{11}, \ldots, y_{1n_1}, \ldots, y_{k1}, \ldots, y_{kn_k})$ the posterior probability of model $\cM_\ell$ is given by
\begin{equation} \label{eq:probbic}
Pr(\cM_\ell \mid Y=y^N) = \frac{Pr(\cM_\ell) \lambda_\ell(y^N)}{\sum_{k=1}^LPr(\cM_k) \lambda_k (y^N) }
\end{equation}
where
$\lambda_{\ell}(y^N):= p(y^N \mid \cM_\ell)= \int_{\Theta_\ell}
\mathcal{L}_N(\cM_\ell, \theta_\ell) p_{\ell}(\theta_\ell) d
\theta_\ell$,
$\ell=1, \ldots, L$ denotes the marginal likelihood (\cite{wass2000}
and \cite{clae2008} among others). Note that the denominator is the
same for every model under consideration, so that we only have to
compare the numerators in \eqref{eq:probbic} in order to compare the
models. Additionally, if we choose equal prior weights for the models,
namely $Pr(\cM_\ell)= 1/L$ for $\ell=1, \ldots, L$, it suffices to
consider the terms $\lambda_1, \ldots, \lambda_L$ for model
selection. In this case, exact Bayesian Inference would use
$2 \log \lambda_{\ell}(y^N)$ for model $\cM_\ell$ to compare between
different models. For the BIC this value is
approximated. Approximating the marginal likelihoods by a Laplace
approximation one obtains (\cite{clae2008})
\begin{equation*}
\lambda_\ell(y^N)\approx \mathcal{L}_N(\cM_\ell, \hat{\theta}_\ell) (2\pi)^{d_{\cM_\ell}/2} N ^{-d_{\cM_\ell}/2} \mid J(\hat{\theta}_\ell) \mid^{-1/2}  p_{\ell}(\hat{\theta}_\ell).
\end{equation*}
Therefore, the approximation is given by
\begin{equation*}
\begin{split}
2 \max_{\theta_\ell} \log(\mathcal{L}_N(\mathcal{M}_\ell, \hat{\theta}_\ell)) - d_{\cM_\ell} \log(N) +d_{\cM_\ell} \log(2\pi) - \log (| J_{\cM_\ell}(\hat{\theta}_{\ell})|)+2 \log(p_{\ell}(\hat{\theta}_{\ell}) ).
\end{split}
\end{equation*}

The penalty term of the BIC only uses the terms of the approximation
which converge to infinity with increasing sample size $N$:
\begin{equation}
  \label{eq:bic}
\BIC(\mathcal{M}_\ell)= 2 \max_{\theta_\ell}\log(\mathcal{L}_N(\mathcal{M}_\ell, \theta_\ell)) - d_{\cM_\ell} \log(N).
\end{equation}
\cite{drap1995} proposed to add the constant term $\log(2\pi)
p_{\ell}$ in (\ref{eq:bic}) and we refer to this
modification of the BIC as $\BIC_2$.

\subsection{Properties}
In this section we investigate two properties for the model selection
criteria introduced so far.  First, we discuss consistency as a method
to compare different model selection criteria (\cite{clae2008}) and
illustrate the theoretical results with a simulation study.  Second,
we investigate the behavior of the criteria if the effect size (the
ratio of treatment effect and variability) stays constant, but the
sample size changes, which is of particular importance when designing
dose finding studies in pharmaceutical drug development.

\subsubsection{Consistency}\label{sec:consistency}

\begin{figure}[t!]
\centering
 \includegraphics[width=0.65 \textwidth]{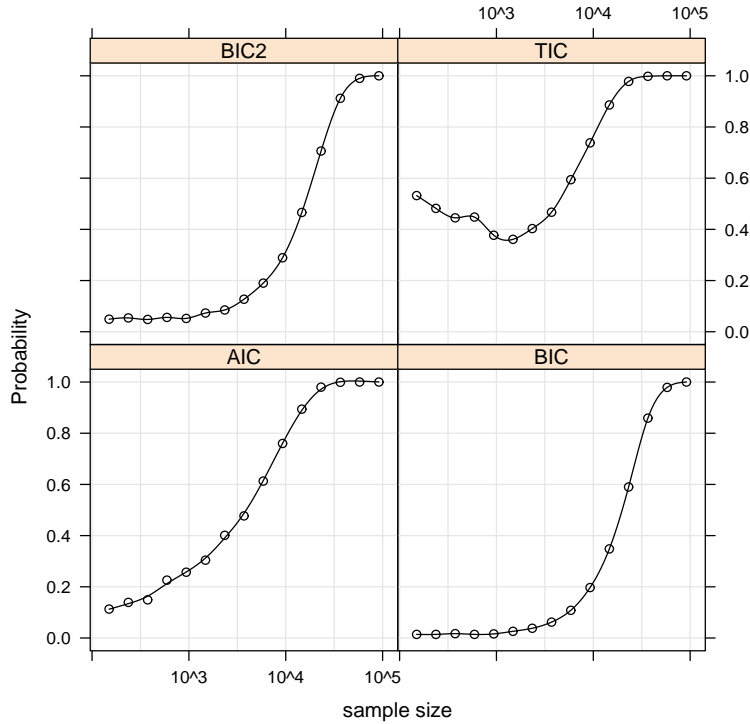}
\caption[Consistency Sigmoid Emax]{\emph{The Probability to select the Sigmoid Emax model if the Sigmoid Emax model is true. The depicted lines are the smoothing splines using the data.}}
\label{fig:consistency_sigemax}
\end{figure}

Consistency is a popular way to compare different model selection
criteria. Consistency of an information criterion ensures that it
picks the best approximating model (among the candidate models) with a
probability converging to 1 with increasing sample size.  In general,
consistency of a model selection criterion of type
\eqref{eq:allcrit} depends on the structure of the penalty term.  If
the best approximating model is unique, a sufficient condition for
consistency is that the penalty term is strictly positive and when
divided by the sample size converges to zero (see \cite{clae2008},
pp. 100-101). All model selection criteria considered in Section
\ref{sec:aic} and \ref{sec:bic} fulfill this requirement. However, if
the best approximating model is not unique and there exist several
best approximating models with different complexities (i.e. nested
models), criteria with a fixed penalty (independent of the sample
size) will not necessarily select the model with the smallest number
of parameters in the set of best approximating models (see
\cite{clae2008}, pp. 101-102). Therefore the $\AIC$ and the TIC have
a tendency to overfit, whereas the $\BIC$ and the $\BIC_2$ do not.

We illustrate this using simulations.  For simplicity, we consider a
situation with two candidate models: Emax and Sigmoid Emax; see
Table~\ref{tab:candmod}.  The Emax model is nested within the Sigmoid
Emax model when setting $\vartheta_3=1$.  We assume a fixed design
where patients are equally randomized to one of the active doses $d_1=
0, d_2=1, \ldots,d_9=8$ and consider increasing sample sizes, starting
with sample size $N=150$, increasing to $N=150,000$.  In the first
scenario the Sigmoid Emax model is the correct model with parameter
$\theta=(0, -1.81, 0.79, 2)$.  As predicted by asymptotic theory the
AIC, the BIC, the $\BIC_2$ and the TIC select the right model with
probability tending to 1, because there is a unique best approximating
model, namely the true Sigmoid Emax model itself (see Figure
\ref{fig:consistency_sigemax}). Comparing the rates of convergence for
the different criteria, we conclude that AIC and TIC perform better
than BIC and $\BIC_2$ in this
scenario.

\begin{figure}[t!]
\centering
\includegraphics[width=0.65 \textwidth]{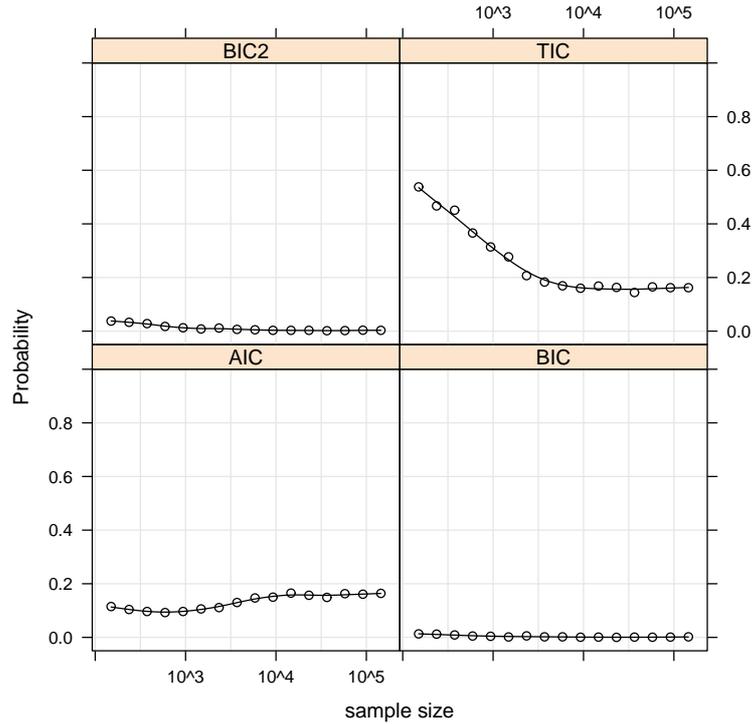}
\caption[Consistency Emax]{\emph{Probability to select the Sigmoid Emax model if the Emax model is true. The depicted lines are the smoothing splines using the data.}}
\label{fig:consistency_emax}
\end{figure}

In the second scenario the Emax model is the true model with parameter
$\theta= (0, -1.81, 0.79)$. Both the Emax and the Sigmoid Emax model
are closest to the true model with respect to the KL-divergence,
because the Emax model is a special case of the Sigmoid Emax model
with $\vartheta_3=1$. As expected the $\BIC$ and $\BIC_2$ choose the
more complex Sigmoid Emax model with probability tending to 0 (see
Figure \ref{fig:consistency_emax}). The AIC and TIC choose the Sigmoid
Emax model with probability tending to $15.7\%$. This value is the
asymptotic probability that the AIC selects the Sigmoid Emax model,
since $\AIC(\mbox{Sigmoid Emax}) - \AIC(\mbox{Emax})
\stackrel{\mathcal{D}}{\longrightarrow} \chi_1^2 -2$ and $P(\chi^2_1 >
2)= 15.7\%$ (see \cite{clae2008}, p. 50).  Summarizing, both the AIC
and the TIC have a tendency to overfit asymptotically if the best
approximating model is not unique.

\subsubsection{Dependence on the sample size}
\label{sec:depend-sample-size}
\begin{figure}[t!]
\centering
 \includegraphics[width=0.65 \textwidth]{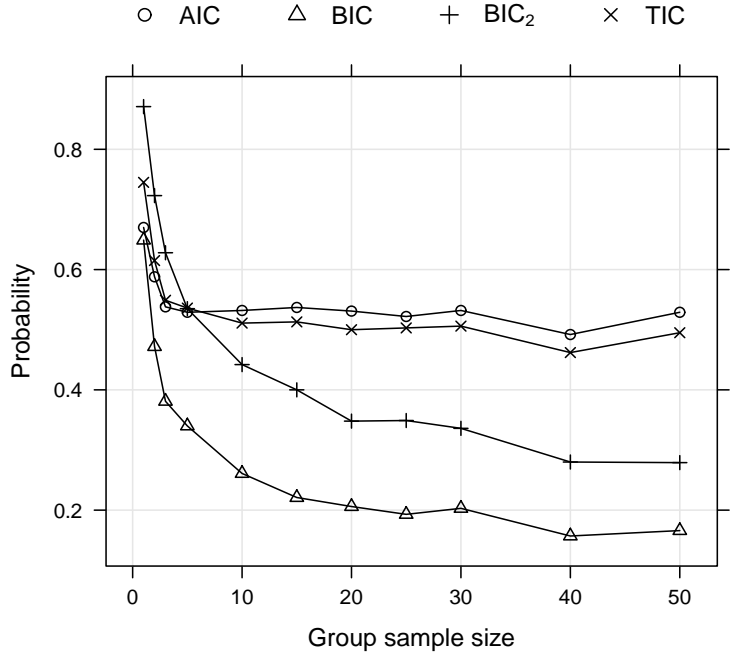}
\caption[Emax Plots varying $\sigma^2$]{\emph{Probability to select the Sigmoid Emax model if the Sigmoid Emax model is true and the variance depends on the group sample size. The depicted lines are the smoothing splines using the data.}}
\label{fig:vary_emax}
\end{figure}

In clinical practice, the sample size at the design stage is often
calculated to ensure that the standard error of a quantity of interest
(typically the treatment effect) is below a given threshold.
From a practical viewpoint it would thus be desirable if a model
selection criterion chooses the same dose response model regardless of
the sample size as long as the standard error around the estimated
dose-response curve remains the same.

In this context one would expect that a model selection criterion
behaves similarly if there are $100$ noisy observations (e.g., with a
standard deviation of $\sigma_y = 10$, thus resulting in a standard
error $\sigma_y/\sqrt{100}= 1$) or if there are $10$ less noisy
observations (e.g., $\sigma_y = \sqrt{10}$ resulting in the same
standard error $\sigma_y /\sqrt{10}=1$).

We investigate the different model selection criteria with respect to
this property using the following scenario. Consider the balanced case
with equal group sample size $n=n_i$ at each dose level $d_i, i=0, 1,
\ldots, 8$. The observations are simulated from the Sigmoid Emax model
with parameter $\theta= (0, -1.81, 0.79, 2)$ and normally distributed
errors with standard deviation $\sigma_n= \sqrt{0.01n}$. That is, the
variance increases with the sample size. Standard results on maximum
likelihood estimation show that the standard error of all estimators
depends on the sample size only through the ratio $\sigma_n /
\sqrt{n}$ and consequently this choice gives a constant standard error
across the
different sample sizes.  The candidate models are again the Emax and the Sigmoid Emax model and the group sample size is given by $n=1, 2, 3, 5, 10, 15, 20, \ldots, 50$. We calculate the probability that the model selection criteria AIC, BIC, $\BIC_2$ and TIC select the Sigmoid Emax model under the assumption that the latter is the true model. The results are displayed in Figure \ref{fig:vary_emax}.\\
We observe that the probability to choose the Sigmoid Emax model is
nearly constant for the AIC and TIC, unless the sample size is very
small. On the other hand the $\BIC$'s and the $\BIC_2$'s probabilities
depend on the sample size since the probability to select the Sigmoid
Emax model decreases with increasing sample size. Thus the sample size
influences model selection by BIC-type criteria not only through the
standard error but also on its own. This is an important point to take
into account when planning studies using BIC-like criteria.

\subsection{Model averaging}\label{sec:mav}
Instead of selecting one model, model averaging can also be
considered. From a Bayesian perspective model averaging arises as soon
as a prior distribution supported on a candidate set of models is
used, because the posterior distribution will then also be based on
the same candidate models, weighted by their posterior model
probability; see \cite{wass2000}. Non-Bayesian model averaging methods
have also been proposed; see \cite{hjor2003} for a detailed
description.  For a given quantity of interest, say $\mu$, these model
averaging estimators are obtained by calculating a weighted average of
the individual estimators of the candidate models $\cM_1, \ldots,
\cM_L$. One way of determining the model weights is to use
transformations of the model selection criteria
for each candidate model. More precisely, let
$\mu$ denote the parameter of interest (e.g. the effect at a specific
dose level) and $\hat{\mu}_\ell$ the estimator of $\mu$ using the model
$\cM_\ell,\ell=1, \ldots, L$. Then the model averaging estimator based on the
model selection criterion $I$ is given by
\begin{equation}\label{eq:mae}
\hat{\mu}_{I, AV}= \sum_{\ell=1}^L \omega_I(\cM_\ell) \hat{\mu}_\ell
\end{equation}
with corresponding weights \cite{hjor2003, buck1997}
\begin{equation}\label{eq:wemae}
\omega_I(\cM_\ell)= \frac{\exp(0.5 \ I(\cM_\ell) )}{\sum_{i=1}^L\exp(0.5 \ I(\cM_i) )}.
\end{equation}
In Section \ref{sec:sim} we will investigate model averaging
estimators for each of the information criteria in Table
\ref{tab:msc}. Note that when $\BIC$ or $\BIC_2$ are used, the
resulting model weights are approximations of the underlying posterior
model probabilities in a Bayesian model, although other criteria, such
as the AIC have a Bayesian interpretation as well; see
\cite{clyd2000}.

\subsection{Bootstrap model averaging based on AIC and BIC}\label{sec:boot}
An alternative way to perform model averaging is to use bagging
(bootstrap aggregating), as proposed by \cite{brei1996}.  In the following
we investigate two estimators of $\mu$ based on bootstrap model averaging using either
AIC or BIC. The essential idea is to bootstrap the model selection
and use all bootstrap predictions for one final prediction. As different
models might have been selected in each bootstrap resample, this method
can also be considered as a model averaging method.

To be precise consider the sample $(d_1, y_{11}),\ldots, (d_1,
y_{1n_1}), \ldots, (d_k, y_{k 1}), \ldots, (d_k, y_{k n_k})$ with
$\sum_{i=1}^k n_i = N$, where $n_i > 1$ for all $i=1, \ldots, k$.  We
determine the bootstrap estimator of $\mu$ based on the AIC and the BIC
using $R$ bootstrap samples:\begin{enumerate}
\item Bootstrap step: \\
Perform a stratified bootstrap on the sample
$$(d_1, y_{11}),\ldots, (d_1, y_{1n_1}), \ldots, (d_k, y_{k 1}) \ldots, (d_k, y_{k n_k}).$$
That is, we select a random sample $(d_i, y^*_{i1}),\ldots,
(d_i, y^*_{in_i})$ of size $n_i$ out of $(d_i, y_{i1}),\ldots, (d_i,
y_{in_i})$ with replacements for every dose $d_i$, $i=1, \ldots, k$.
\item Model selection step: \\
Calculate the AIC and BIC value for every competing model based on the bootstrap sample. Select the model with the largest AIC and BIC value and estimate the parameter of interest $\mu$ based on the selected model. The resulting estimators are  denoted by $\hat{\mu}^B_{\AIC}$ and $\hat{\mu}^B_{\BIC}$, respectively.
\end{enumerate}

From the $R$ bootstrap samples the medians of the $R$ different
estimators $\hat{\mu}^B_{\AIC}$ and $\hat{\mu}^B_{\BIC}$ are used as the
bootstrap estimators for $\mu$.

\section{Simulations}\label{sec:sim}
In this section, we report the results of an extensive simulation to
investigate and compare the different model selection and averaging
approaches in scenarios that are realistic for Phase II dose finding
trials.  In Section~\ref{ssec:des} we introduce the design of the
simulation study, including its assumptions and scenarios.  In
Section~\ref{ssec:measure} we describe the performance measurements
used to evaluate the different approaches.  In
Section~\ref{ssec:sim} we summarize the results of the simulation
study.

\subsection{Design of simulation study}
\label{ssec:des}
Following the simulation setup of \cite{phrma:2007}, we investigate
different constellations of
sample sizes, number of active dose levels, and true dose response models.
More precisely, we consider two sample sizes $N=150$ and $N=250$ for each of four different designs $\Xi=(d_1, \ldots, d_k)$
of $k=k(\Xi)$ active dose levels,
assuming either five ($A=\{0,2,4,6,8\}$, $k(A)=5$), seven
($B=\{0,2,3,4,5,6,8\}$, $k(B)=7$), nine ($C=\{0,1,2,3,4,5,6,7,8\}$,
$k(C)=9$) or four ($D= \{0,2,4,8\}$, $k(D)=4$) active dose levels.  In
each case the total sample size $N$ is equally distributed across the
different active dose levels. If $\tfrac{N}{k(\Xi)}$ is not an
integer, we use a rounding procedure provided by
\cite[p. 307]{pukelsheim2006}.  Further, we assume the five dose
response models described in Section \ref{sec:case} with parameters
given in Table \ref{tab:candmod} as true models in the simulation.
The errors in model \eqref{eq:model} are normally distributed with standard deviation $\sqrt{4.5}$.\\
Thus, a scenario $S$ is defined by the used total sample size $N$, the
used design $\Xi$ and the model used for generating the data. For
example, one scenario is given by $N=150$, $\Xi=C$ and the Emax model
as data generating model. Summarizing, there are $\mathcal{S}= 2 \cdot
4 \cdot 5 = 40$ scenarios (two possible sample sizes, four possible
designs and five possible data generating models).

In the first simulation study we exclude the ANOVA model from the list
of candidate models under consideration, focussing on the first four
models in Table \ref{tab:candmod}. That is, if the ANOVA model is used
for generating the data, no dose response model in the candidate set
can exactly fit the underlying truth, so that in this case we
investigate the behavior under model misspecification. In the second
simulation study we will add the ANOVA model to the candidate models
under consideration, thus using all five models from Table
\ref{tab:candmod}.  Furthermore, we exclude the Sigmoid Emax model
from the set of candidate models in scenarios based on the design $D$,
since its parameters are not estimable under $D$.  All results are
based on $N_{sim}= 1000$ simulation runs per scenario. In each
simulation run the parameters of the different candidate models are
estimated and the value $I(\mathcal{M})$ is calculated for each model
selection criterion specified in Table \ref{tab:msc} and for each dose
response model specified in Table \ref{tab:candmod}.  The bootstrap
model averaging approach was used with $R=500$ bootstrap simulations
for each simulated trial.\\
All simulations are performed using the R-package DoseFinding
\cite{born2013}.

\subsection{Measurements of performance }
\label{ssec:measure}
We use the standardized mean squared error (SMSE) and the averaged standardized mean square error (ASMSE) to assess estimation
of the dose effects and the target dose of interest, as well the
proportion of selecting the correct dose response model to evaluate
the performance of the model selection criteria.

For a given scenario $S$ (out of the $\mathcal{S}=40$ scenarios) let $\hat{\eta}_{j, S}(d, \hat{\theta}_{j, S})$ denote the estimated regression
model with corresponding estimated model parameters $\hat{\theta}_{j, S}$ which is
selected by a given model selection criterion $I$ in the $j$-th simulation
run. Moreover, let $\eta_S(\cdot, \theta_S)$ denote the data generating dose response
model of the scenario $S$.  The mean squared error (MSE) of the treatment
effect estimator at dose level $d$ is then given by
\begin{equation*}\label{eq:mse_de}
\mse(d, S)=\frac{1}{N_{sim}} \sum_{j=1}^{N_{sim}} \left(\hat{\eta}_{j, S}(d, \hat{\theta}_{j, S})- \eta_S(d, \theta_S)\right)^2.
\end{equation*}
The average mean squared error (AMSE) for an arbitrary design $\Xi$ with $k(\Xi)$ different active dose levels is given by
$\amse(\Xi, S)=\frac{1}{k(\Xi)} \sum_{i=1}^{k(\Xi)} \mse(d_i, S)$.
In order to obtain comparability between the scenarios it is useful to standardize the average mean squared errors. This is achieved by
dividing $\amse(\Xi, S)$ by the minimal average mean squared
error $\mmse(\Xi, S) = \min_{\eta_{\cM}} \frac{1}{N_{sim}} \sum_{j=1}^{N_{sim}} \frac{1}{k(\Xi)} \sum_{i=1}^{ k(\Xi)} \bigl(\eta_{\cM}(d_i, \hat{\theta}^{(k)}_{\cM}) - \eta_S(d_i, \theta_S) \bigr)^2$
where the minimum is taken with respect to all models $\eta_{\cM}$ and $\hat{\theta}^{(k)}_{\cM}$ is the maximum likelihood estimator of model $\cM$ in the $k$-th simulation run of scenario $S$.
The standardized MSE (SMSE) of scenario $S$ for a specific selection
criterion is then given by
\begin{equation}\label{eq:smse_xi}
\mbox{SMSE}(\Xi, S) = \frac{\amse(\Xi,
  S)}{\mmse(\Xi, S)}
 \end{equation}
 and the averaged standardized MSE (ASMSE), \textit{i.e.} the SMSE
 averaged over all simulation scenarios, is given by
\begin{equation}\label{eq:rmse}
\mbox{ASMSE}(\Xi) = \frac{1}{\mathcal{S}} \sum_{s=1}^\mathcal{S} \mbox{SMSE}(\Xi, S_s).
\end{equation}

Moreover, we consider model selection procedures to estimate the
target dose achieving an effect of $\delta=1.3$ over placebo
$\td_{\eta_{\cM}}( \theta_{\cM})= \eta^{-1}_{\cM}(\delta, \theta_{\cM})$
for a given dose response model $\cM$.
Similarly as above, we then define the MSE as
\begin{equation}\label{eq:mse_td}
\mse_{\td, \eta}( S)=\frac{1}{N_{sim}} \sum_{j=1}^{N_{sim}} \left({\td}_{\hat{\eta}_{S,k}}(\hat{\theta}_{S,k}) - \td_{\eta_S}( \theta_S)\right)^2.
\end{equation}
Note that those simulation runs, where the estimated target dose is not contained within the dose range, are excluded from the MSE calculation.
The standardization of the MSE is again achieved by dividing \eqref{eq:mse_td} by $\mmse_{\td}(S)=\min_{\eta_{\cM}} \frac{1}{N_{sim}} \sum_{k=1}^{N_{sim}} \bigl(\td_{\eta_{\cM}}(\hat{\theta}^{(k)}_{\cM}) - \td_{\eta_S}(\theta_S)\bigr)^2$
where the minimum is calculated with respect to all models
$\eta_{\cM}$ under consideration. The standardized MSE (SMSE) of scenario $S$ for the target dose is then given by
\begin{equation}\label{smse_td}
\mbox{SMSE}_{\td}( S) = \frac{\mse_{\td}(S)}{\mmse_{\td}( S)}
 \end{equation}
and the averaged standardized mean square error for the target dose ($\mbox{ASMSE}_{\td}$) is again obtained by averaging the SMSEs over all scenarios.\\
Note that the estimator of the target dose is calculated by interpolation if the ANOVA model is selected by the model selection criterion $I$.

\medskip
\noindent
The model averaging estimators $\hat{\mu}_k(d)$ for the dose effect $d$
and the target dose are obtained from \eqref{eq:mae} and
\eqref{eq:wemae}, where the parameter $\mu$ is given by $\eta(d,
\theta)$ and $\td_{\eta}(\theta)$, respectively. The definition of
the weights in the model averaging procedure is slightly modified if
the target dose estimator of a model lies outside the dose range. In
this case the estimator is not used and the model averaging estimator is
calculated from the weights of the remaining models if their weights
sum up to a value greater than $20\%$. Otherwise this case is excluded.
For bootstrap model averaging a similar approach is used, when there
are more than $80\%$ of the target dose estimators lying outside the
design space for a given bootstrap run, it is excluded from the
calculation.

\subsection{Simulation Results}\label{ssec:sim}

In Section \ref{sec:genres} and Section \ref{sec:ms vs mav} we present
the simulation results corresponding to the candidate set consisting of linear, quadratic, Emax and Sigmoid Emax model. In
\ref{sec:ancand} we analyze how the performance of the model selection
criteria and model averaging methods change if the ANOVA model is
added to the candidate set.

\subsubsection{Results based on the candidate models 1-4 in Table \ref{tab:candmod}}\label{sec:genres}

First, we consider the case where the ANOVA model is not among the
candidate models used for analysis. In Table \ref{tab:averages_ms} and
\ref{tab:averages_mav} we display the ASMSEs defined in \eqref{eq:rmse}
for the designs $A, B, C, D$ and for the target dose.

\begin{table}[t!]
\centering
\begin{tabular}{|l | ccccc|}
\hline
  criterion  &       AIC  &     BIC  &    $\BIC_2$ &       TIC &    $\AIC_C$ \\
\hline
ASMSE(A) &  \textbf{1.35} & 1.54 & 1.41 & 1.35 &1.51 \\ 
ASMSE(B)  & \textbf{1.38} &1.56&1.44 &1.38 & 1.55\\ 
ASMSE(C)  & \textbf{1.35} & 1.46 & 1.38 & 1.35 & 1.48\\
ASMSE(D)  & \textbf{1.37}& 1.58 & 1.44 &1.37 &1.55\\ 
$\mbox{ASMSE}_{\mbox{td}}$ & \textbf{1.70} &2.35 &1.92 & 1.71 &2.74 \\ 
\hline
\end{tabular}
\caption[Averages MAV]{\textit{The averaged standardized mean squared
    errors ({\rm ASMSE}, cf. \eqref{eq:rmse}) for the designs A, B, C
    and D and for the target dose
    under the consideration of different model selection
    criteria. (The best values per row are printed in bold.)
  }}
\label{tab:averages_ms}
\end{table}

\begin{table}[t!]
\centering
\begin{tabular}{|lccccccc|}
\hline
 criterion      &    AIC   &   BIC    & $\BIC_2$    &  TIC  &   $\AIC_C$ & AIC-Boot & BIC-Boot \\
\hline
ASMSE(A) & \textbf{1.24} & 1.39 & 1.28& 1.24 & 1.26 &  \textbf{1.21} & 1.29 \\

ASMSE(B)  & \textbf{1.26 } & 1.40 & 1.30 & 1.26 & 1.28 &  \textbf{1.24} & 1.30\\

ASMSE(C)&  \textbf{1.23} & 1.33 & 1.25 & 1.24 & 1.25 &  \textbf{1.21} & 1.24 \\

ASMSE(D)  & \textbf{1.25} & 1.42 & 1.30 & 1.25 & 1.27&  \textbf{1.23} & 1.31\\

$\mbox{ASMSE}_{\mbox{td}}$ & \textbf{1.54} & 1.88 & 1.62 & 1.54 & 1.90 &  \textbf{1.30} & 1.44\\
\hline
\end{tabular}
\caption[Averages MAV]{\textit{The averaged standardized mean squared errors ({\rm ASMSE}, cf. \eqref{eq:rmse}) for the designs A, B, C and D and for the target dose with respect to model averaging and bootstrap model averaging. (The best values per row are printed in bold.)}}
\label{tab:averages_mav}
\end{table}

\begin{figure}[t!]
\subfigure{\includegraphics[width=0.5 \textwidth]{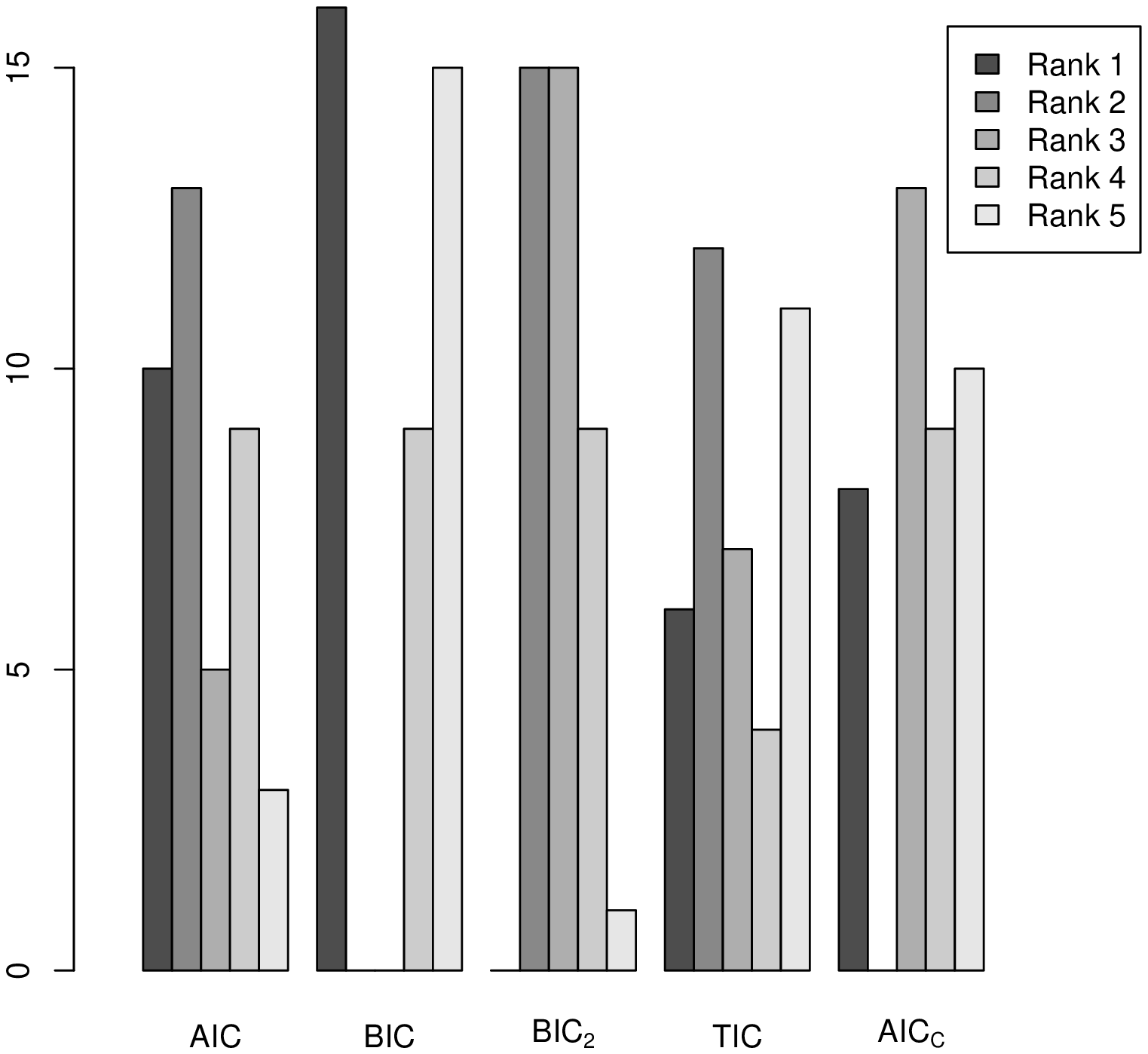}}
\subfigure{\includegraphics[width=0.5 \textwidth]{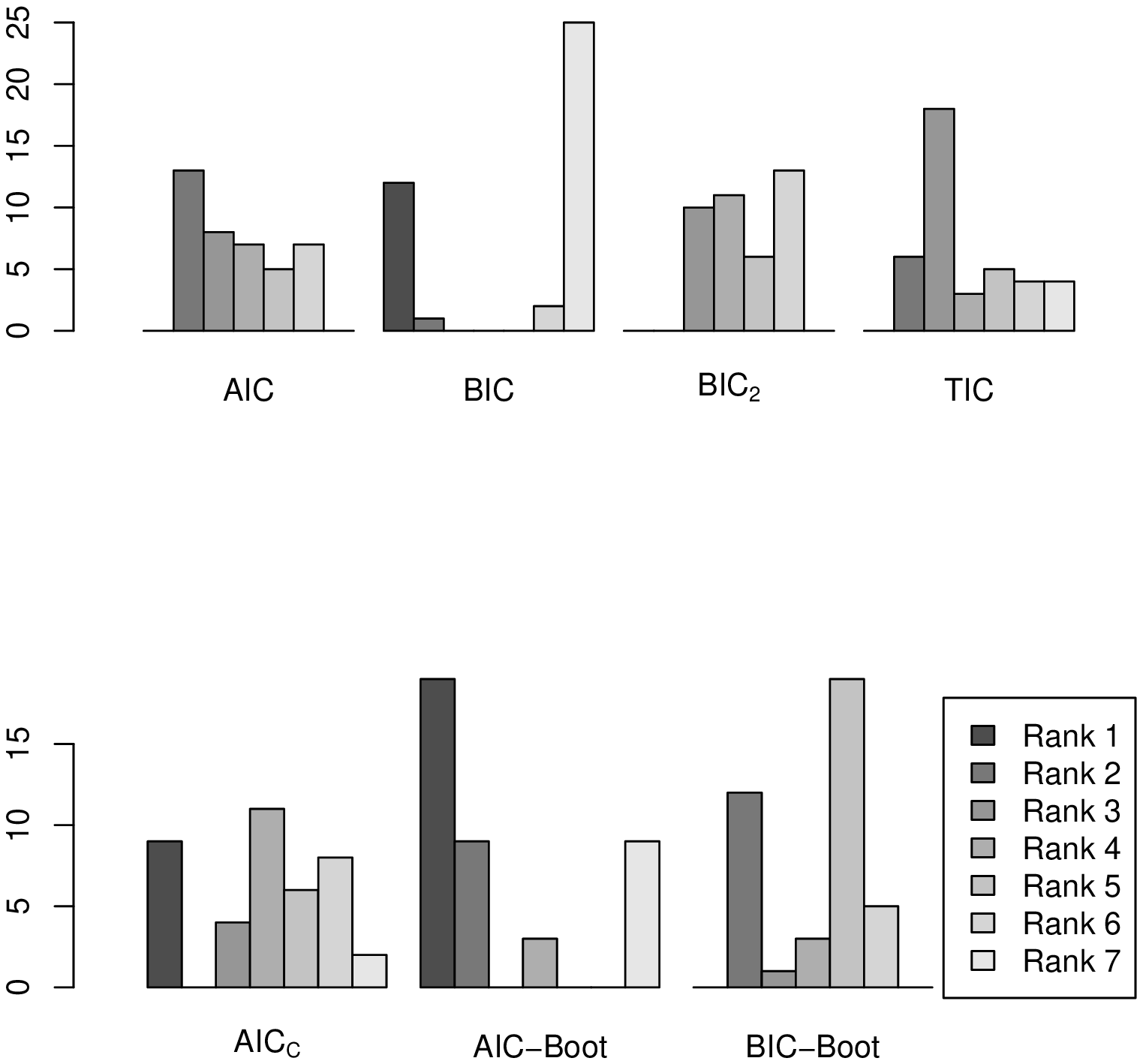}}
\caption[Rang MS MSEC]{\emph{The distribution of the ranks over all
    scenarios for the SMSE of dose effect in design $\Xi=C$. Left:
    model selection, right: model averaging methods.}}
\label{fig:rang_msmav_msec}
\end{figure}

The AIC and the TIC perform similarly and have the best average
performance across all scenarios both for model selection and model
averaging. Comparing model selection and model averaging it can be
observed that the model averaging procedures generally perform
slightly better on average than model selection.

To get an idea of the performance in each individual scenario we
ranked the model selection and model averaging approaches for each
scenario according to their performance and display the ranks in
Figure \ref{fig:rang_msmav_msec}. One can clearly see that almost all
criteria perform best in some scenarios and worst in other scenarios,
so no clear best criterion can be identified. It is interesting to
observe, however, that for the BIC the performance is either very good
or very bad, while for the AIC or TIC the performance is more balanced
across all scenarios. The mixed performance of the BIC is due to the
fact that it penalizes the complexity of a model more strongly. Consequently, it prefers the linear model because of its
smaller number of parameters, even if it is not an adequate
model. However, in situations when the linear model is the true one,
the BIC performs best.

In terms of probabilities to select the true model, we observe that
the AIC performs best with respect to these criteria; see
Figures~\ref{fig:auswahl_aic_aicc}, \ref{fig:auswahl_bic_bic2}, and
\ref{fig:auswahl_tic} in Appendix \ref{sec:prob} for the detailed
results.  The averaged probabilities over all scenarios
 to select the true model are given by $43\%, \,
34\%, \, 39\%, \, 42\% \mbox{ and} \, 23\% $ for the
AIC, BIC, $\BIC_2$, TIC and $\AIC_C$, respectively, which also show
some advantages for model selection based on the AIC.  Summarizing,
with this set of candidate models, (linear, quadratic, Emax and
Sigmoid Emax), the AIC based estimators AIC and TIC (both for model
selection as well as for model averaging) outperform those based the
other criteria.


\subsubsection{Model Selection vs. Model Averaging}\label{sec:ms vs mav}
\begin{figure}[t!]
\begin{center}
\includegraphics[width=0.7\textwidth]{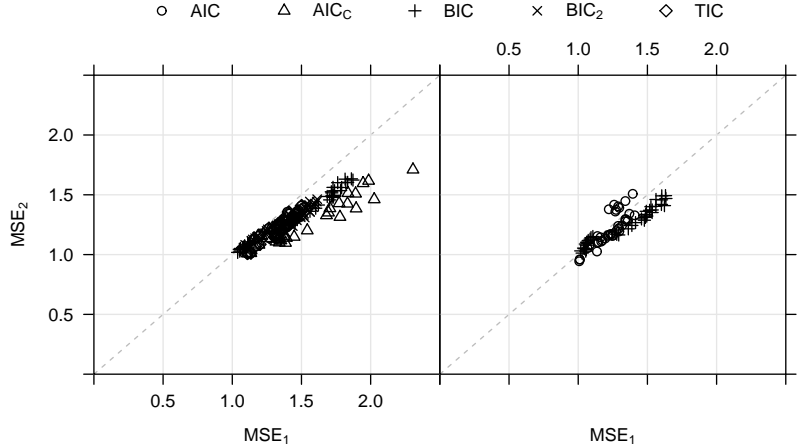}
\end{center}
    \caption[Comparison Dose Response]{\emph{Comparison of model
        selection, model averaging and bootstrap for estimating the
        dose effects in design $C$. The Figure shows the $\mbox{SMSE}$
        values. Left panel: model selection ($\mbox{MSE}_1$) versus model
        averaging ($\mbox{MSE}_2$) . Right panel: model averaging ($\mbox{MSE}_1$) versus bootstrap model
        averaging ($\mbox{MSE}_2$).}}
    \label{fig:comp_de_mse}
\end{figure}

\begin{figure}[t!]
\begin{center}
  \includegraphics[width=0.7\textwidth]{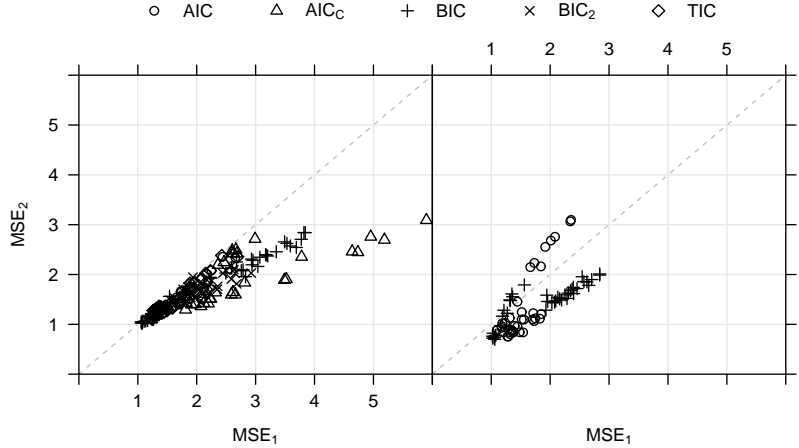}
  \end{center}
   \caption[Comparison Target Dose]{\emph{Comparison of model
       selection, model averaging and bootstrap model averaging for estimating the
       target dose. The Figure shows the $\mbox{SMSE}$. Left panel:
       model selection ($\mbox{MSE}_1$) versus model averaging  ($\mbox{MSE}_2$) . Right panel: weights
       based model averaging ($\mbox{MSE}_1$) versus bootstrap model averaging  ($\mbox{MSE}_2$) .}}
    \label{fig:comp_td_mse}
\end{figure}

In this section we compare the results of model averaging with those
of model selection in more detail. In terms of the average
performance, we observed that model averaging outperforms model
selection (see Tables \ref{tab:averages_ms},
\ref{tab:averages_mav}). We now investigate the individual results for
each scenario, see the left plots in Figures~\ref{fig:comp_de_mse}
and~\ref{fig:comp_td_mse} which correspond to the SMSE of the dose
effect in \eqref{eq:smse_xi} and the target dose in \eqref{smse_td}.
The dashed line in the left panel of Figure~\ref{fig:comp_de_mse}
displays the situations where the SMSE of the model selection based
estimator and the SMSE of the model averaging estimator are equal.
The points below (above) the diagonal correspond to scenarios where
the model averaging (selection) estimators have a smaller SMSE.  For
example, $\mbox{SMSE}(C, S) = 1.73$ for BIC model selection, but
$ 1.48$ for BIC model averaging in the Emax scenario $S$ with sample
size $250$ under design $B$, indicating that the BIC model averaging
estimator is more precise than the model selection estimator in this
scenario.

One observes that across all scenarios model averaging
tends to outperform model selection consistently, resulting in smaller SMSE, even
though the differences are never substantial. The ratio of the $\mbox{SMSE}(C, S) $
for BIC model selection and the $\mbox{SMSE}(C, S) $ or BIC model averaging is
given by $1.17$, which means that on average $17\%$ more observations
are needed for model selection in order to result in an estimator of
similar precision as obtained with the corresponding model averaging
approach.  Note that the individual improvement obtained by model
averaging depends on the selection criterion. Comparing the
improvement by model averaging with respect to target dose estimation
is even more substantial (see the left panel in Figure
\ref{fig:comp_td_mse}).

In the right panels of Figure \ref{fig:comp_de_mse} and of Figure
\ref{fig:comp_td_mse} we compare the model averaging using bootstrap
with that based on the AIC and the BIC weights. We observe that the
bootstrapping estimators yields slightly better results than the model
averaging estimators except for the linear scenarios. For example, the
$\mbox{SMSE}(C, S)$ belonging to BIC model averaging is equal to $1.48$
whereas the corresponding $\mbox{SMSE}(C, S)$ of the BIC bootstrap
estimator is smaller ($\mbox{SMSE}(C, S)=1.29$) in the Emax scenario with
sample size $250$, design $B$ (see red line in the right panel of
Figure \ref{fig:comp_de_mse}).

The reason why the performance is worse for the linear model is that
in general, more complex models are preferred by bootstrap model
averaging (especially when using the AIC) which implicates a lower
selection probability for the linear model. This behavior improves the
performance of the bootstrap estimators in the non linear scenarios
whereas it gets worse in the linear scenarios.

\subsubsection{Simulation Results based on the candidate models 1-5 in Table \ref{tab:candmod}}\label{sec:ancand}
From a practical point of view adding the ANOVA model to the set of
candidate models can be considered helpful to safeguard against
unexpected shapes, as the ANOVA model is extremely flexible.

\begin{figure}[t!]
\centering
\subfigure{\includegraphics[width=0.48 \textwidth]{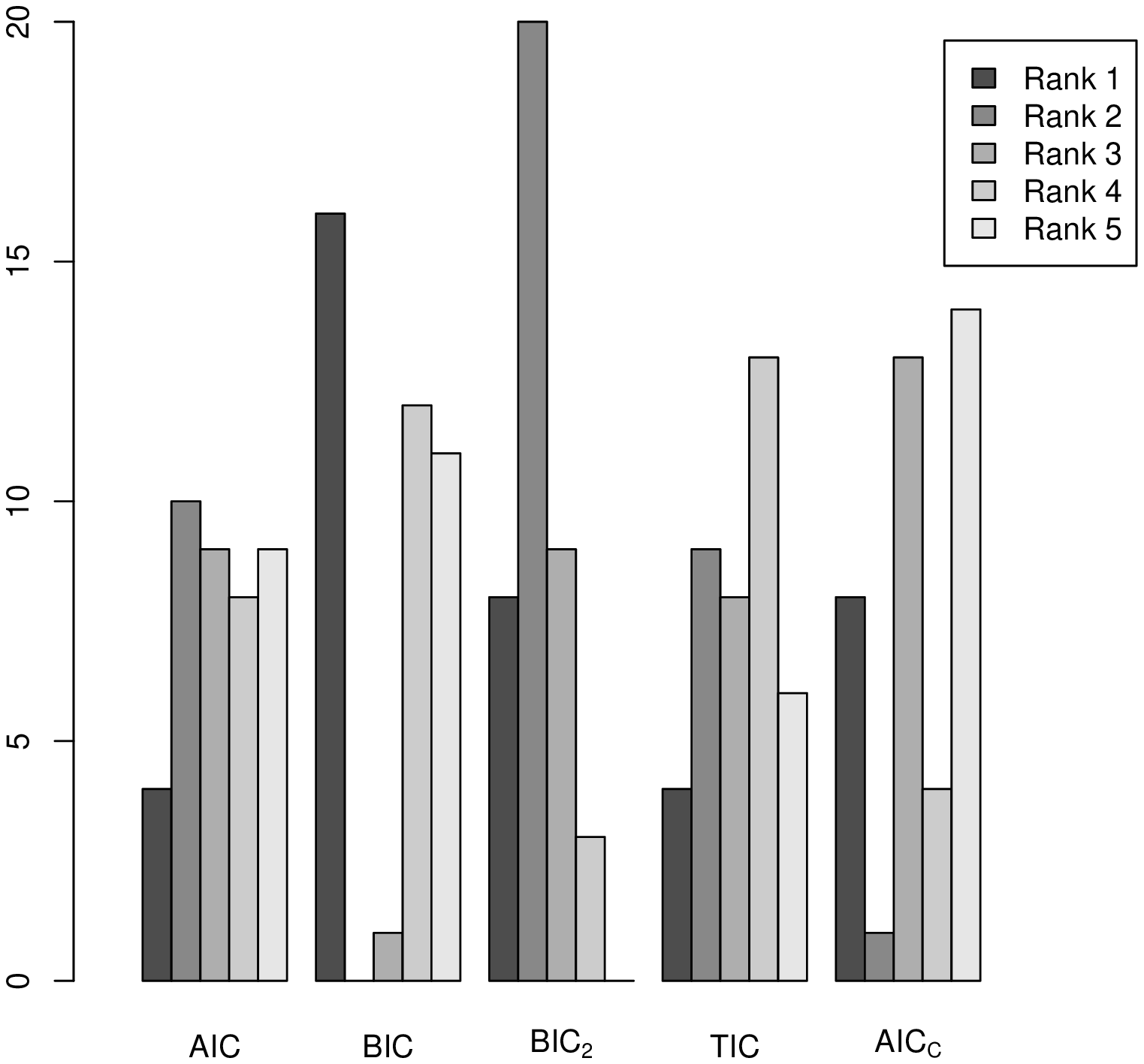}}
\subfigure{\includegraphics[width=0.48 \textwidth]{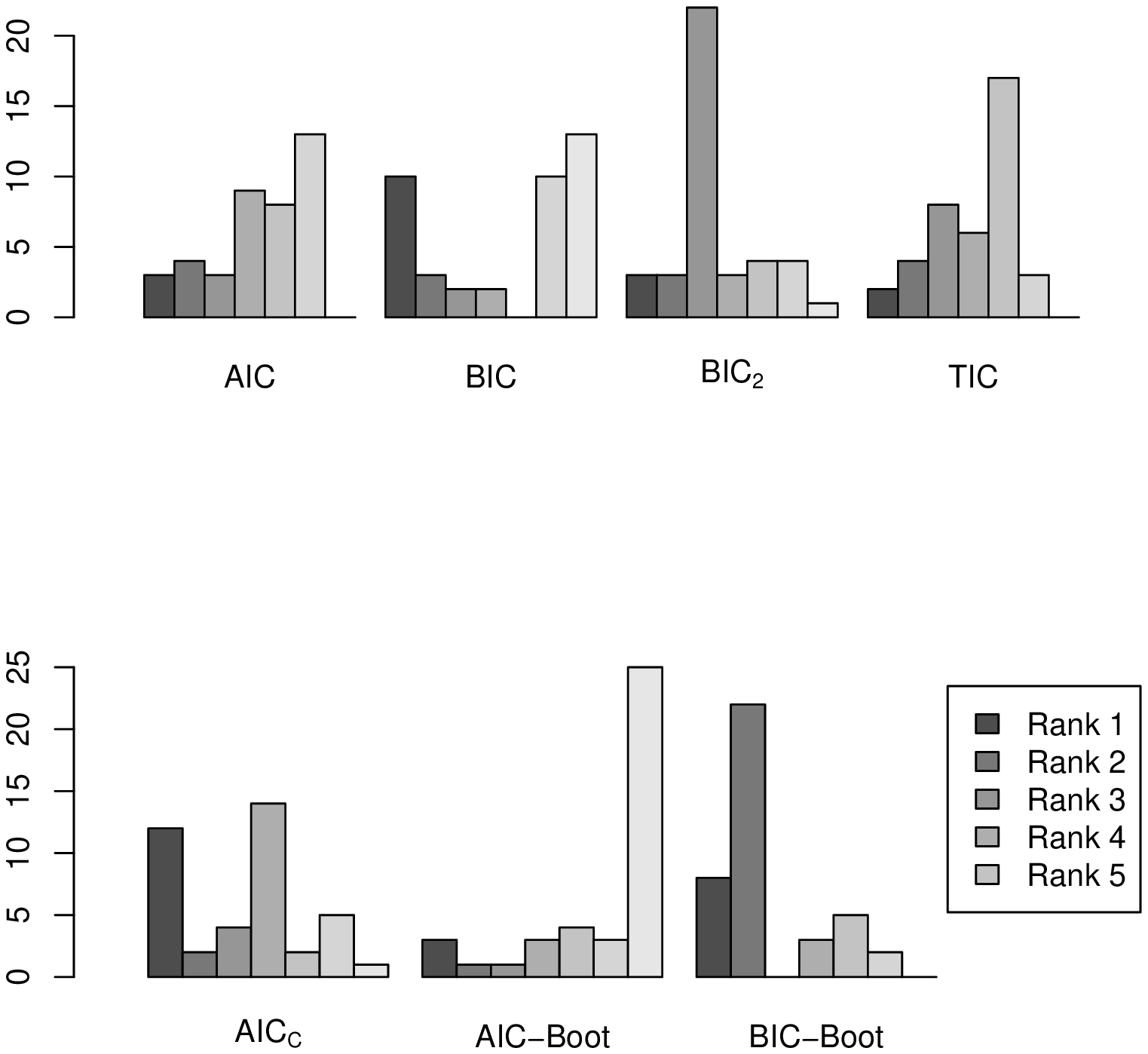}}
\caption[Rang MS MSEC]{\emph{The distribution of the ranks over all scenarios for the metric SMSE dose effect in design C (left: model selection, right: model averaging). The ANOVA model is among the candidate models.}}
\label{fig:rang_ms_msecwa}
\end{figure}

To compare the different criteria for model selection and model
averaging, we calculated the same metrics as in the last section. In
this case the superiority of the AIC and TIC cannot be observed
anymore (see Figure \ref{fig:rang_ms_msecwa}).  The model selection
criteria perform more similarly compared to each other (see Tables
\ref{tab:averages_ms_wA}, \ref{tab:averages_mav_wA} in Appendix
\ref{sec:withanova}). In general, however, model averaging still
outperforms model selection on average, the only exception to this is
bootstrap model averaging based on the AIC. The ANOVA approach
represents a rather complex model (one parameter per dose) and it
seems that the AIC does not penalize this complexity strongly enough,
thus leading to an inferior performance. The BIC is not affected
similarly since it uses a higher penalty.

\begin{figure}[t!]
\centering
\includegraphics[width=.8 \textwidth]{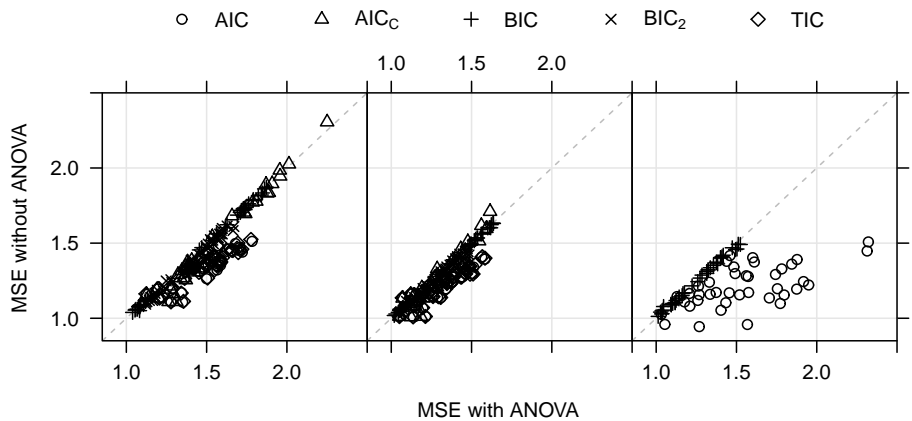}
\caption[Comp nA vs. wA]{\emph{Comparison of the SMSE of dose effect
    estimators in design C with and without the ANOVA model (left
    panel: model selection, center panel: weight based model
    averaging, right panel: bootstrap model averaging).}}
\label{fig:vgl_na_wa}
\end{figure}

Considering the direct comparison of both candidate sets (namely the
one with ANOVA and the one without ANOVA) using the SMSE (see Figure
\ref{fig:vgl_na_wa}, left plot) the criteria mostly perform better if
the ANOVA model is not among the candidate models.

Model averaging estimators also perform better if the ANOVA model is
not among the candidate models. For bootstrap model averaging (see
right panel in Figure \ref{fig:vgl_na_wa}) one can clearly see that
the AIC with the ANOVA candidate model gets much worse, while the BIC
is not affected.

Summarizing, the performance of the model selection criteria
depends sensitively on the candidate model set. Including the ANOVA model does
not improve the performance of all criteria, it sometimes even
deteriorates the performance. Of course this is due to the fact that
the dose response shape for the ANOVA model considered here can be
approximated roughly by the other candidate models. If more extreme,
irregular shapes were used, inclusion of the ANOVA model could improve
performance.

\section{COPD Case Study Revisited}\label{sec:caserev}

Taking into account the results from Sections \ref{sec:not} and
\ref{sec:sim} we now return to the COPD case study and the three
questions posed in Section \ref{sec:case}, which were (i) which of the
candidate models should be used for the dose response modeling step,
(ii) whether model selection or averaging should be used, and (iii)
which specific information criteria should be employed to perform
either model selection or averaging.

All dose response models introduced in Table~\ref{tab:candmod} were
fitted to the COPD data from Section \ref{sec:case}. The model fits
are displayed in Figure~\ref{fig:fitmod} in Appendix
\ref{sec:modfitex}. Visually all model fits are adequate, perhaps with
the exception of the linear model, which seems to overestimate the
placebo response.

\begin{table}[t!]

\centering
\begin{tabular}{|l|rr|rr|rr|}
\hline
\multirow{2}*{candidate models} & \multicolumn{3}{|c|}{AIC} &    \multicolumn{3}{|c|}{BIC}       \\
\cline{2-7}
& values & \multicolumn{1}{r}{weights} &\multicolumn{1}{c|}{Bootstrap} & \multicolumn{1}{r}{values} & weights & Bootstrap \\
\hline
Linear  &52.17 &\multicolumn{1}{r}{15\%} & \multicolumn{1}{r|}{26\%}   &  \multicolumn{1}{r}{41.48} & 58\% & 64\% \\
Quadratic & 53.50& \multicolumn{1}{r}{30\%} & \multicolumn{1}{r|}{36\%}&   \multicolumn{1}{r}{39.24}  & 19\% & 17\% \\
Emax  &  53.84 &\multicolumn{1}{r}{36\%} &\multicolumn{1}{r|}{30\%} &  \multicolumn{1}{r}{39.58} &22\% & 19\% \\
SigEmax & 51.85  &\multicolumn{1}{r}{13\%}&\multicolumn{1}{r|}{0\%} &  \multicolumn{1}{r}{34.03}& 1\%& 0\%\\
ANOVA &     50.14 & \multicolumn{1}{r}{6\%} & \multicolumn{1}{r|}{8\%} & \multicolumn{1}{r}{28.75}& 0\% & 0\% \\
\hline
\hline
\multirow{2}*{candidate models}  &     \multicolumn{2}{|c|}{$\BIC_2$} & \multicolumn{2}{|c|}{TIC}   &  \multicolumn{2}{|c|}{$\AIC_C$}  \\
\cline{2-7}
& values & weights  & values & weights & values & weights \\
\hline
\hline
Linear  & 47.00 & 34\%& 52.40 & 16\% & 52.09 &16\%\\
Quadratic & 46.59 & 28\%  & 53.73&  30\% & 53.36  & 30\%\\
Emax    & 46.93 & 33\% & 54.05  &  35\%& 53.70 & 36\% \\
SigEmax &43.21 & 5\%& 52.07 & 13\% &51.64 & 13\%\\
ANOVA & 39.78 & 0\% &  50.36 & 6\% & 49.85 & 5\%\\
\hline
\end{tabular}
\caption[Example]{\textit{The different values of the selection criteria, the corresponding model averaging weights (in \%) and the relative frequency (in \%) of the AIC and BIC bootstrap in the COPD case study.}}
\label{tab:ms_example}
\end{table}

\begin{figure}[t!]
\centering
\includegraphics[width=.8\textwidth]{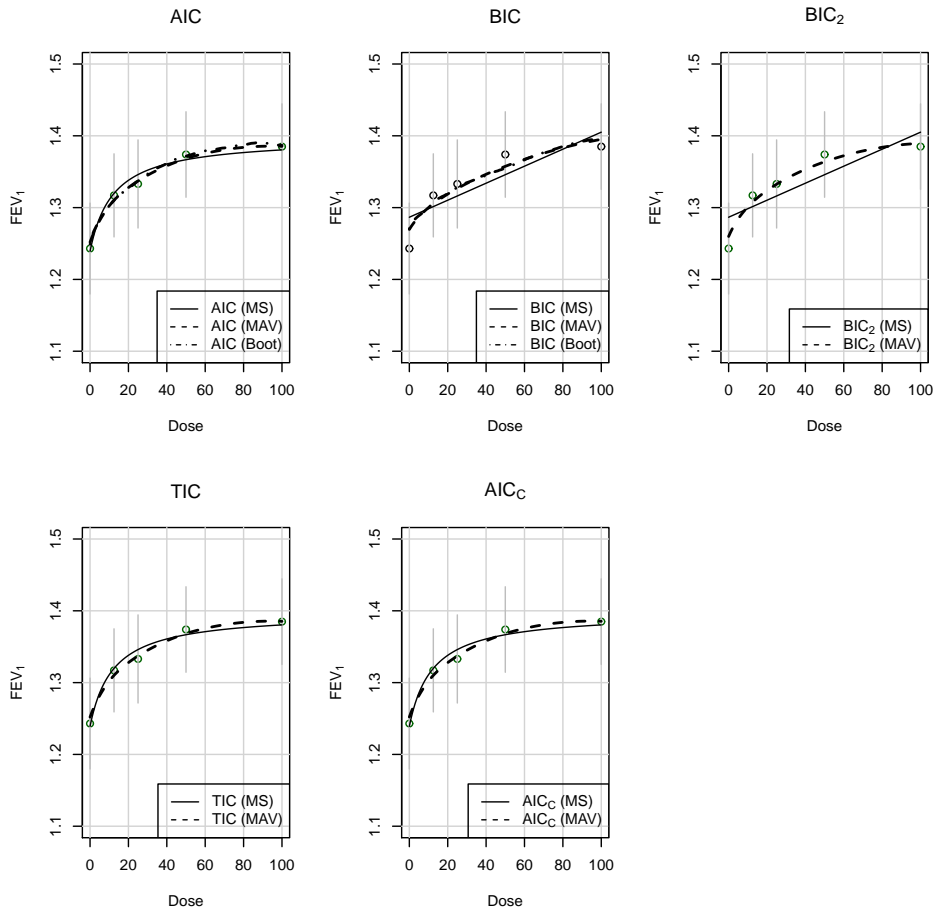}
\caption[Example]{\emph{The fitted models after model selection with respect to different criteria.}}
\label{fig:plots_example}
\end{figure}

When observing the results for the different information criteria in
Table~\ref{tab:ms_example} one can see that the AIC-type criteria are
rather consistent among each other and favor the Emax with~36\%, the
quadratic model with~30\%, followed by the Sigmoid Emax and the
linear model with roughly~15\% each and the ANOVA model with~6\% in
terms of model weights. The BIC-related criteria give more weight to
the linear model (as already observed in the simulations), as they
penalize the number of parameters more strongly. The BIC penalizes
here considerably more strongly than the BIC$_2$, giving 58\% weight to the
linear model, while the BIC$_2$ gives around~30\% to each of the linear,
quadratic and Emax models. The fitted curves based on model selection
and model averaging for all approaches are displayed in Figure
\ref{fig:plots_example}. It can be seen that for most methods the difference between
model selection and model averaging is not very large in this example,
because the models that accumulate model weights lead to relatively
similar fits. For the BIC and the BIC$_2$, however a substantial difference can be
observed between model averaging and selection: The linear model gets
selected, but the Emax and quadratic model have almost equally large
model weights. Model averaging seems particularly important in these
situations, to adequately reflect the uncertainty in the modeling
process.

Regarding question (i) and (ii): The simulations in the last section
showed a consistent benefit of model averaging over selection. So the
proposal would be to use model averaging here. There does not
seem to be a major difference between the weight-based and bootstrap
model averaging in this particular example (see Table
\ref{tab:ms_example} and Figure \ref{fig:plots_example}). Regarding
question (iii), in the simulations for a similar candidate set of
models (which did not include ANOVA) the BIC showed a slightly worse
behavior than the AIC, suggesting the use of the AIC-related criteria in this scenario.

The two main objectives of the study, were to evaluate the shape of
the dose-response curve and doses achieving an effect of $0.1-0.14$
liters on top of placebo. So in our situation we will use model
averaging with the AIC with bootstrapping to answer
these questions. The placebo effect of the curve was estimated with 1.25
(95~\%~CI: [1.18,~1.31]). Within the observed dose-range increases
monotonically up to an effect of 0.14  (95~\%~CI: [0.06,~0.22]) at the maximum
dose of 100mg. At the 50mg dose 0.13 (95~\%~CI: [0.04,~0.20]) 91.22 \% (95~\%~CI: [ 50.00\%,~164.21\%]) of the
maximum effect is achieved, indicating that a plateau-like level is
achieved there. The increasing part of the curve is between 0 and
50mg.


\section{Conclusions}\label{sec:concl}

This paper compared different existing methods for model selection and
model averaging in terms of their mathematical properties and their
performance for dose-response curve estimation in a large scale
simulation study.

In terms of their mathematical properties, it was reviewed and
illustrated that the BIC-type criteria are consistent, while the
AIC-type criteria asymptotically tend to prefer too complex models
(see Figures \ref{fig:consistency_sigemax} and
\ref{fig:consistency_emax}).  It was also investigated, that BIC-type
criteria select different models for different total sample size, even
when the estimated dose-response curves and the uncertainty around
each dose-response curve are the same (\textit{i.e.}  the confidence
intervals width around the curve). This is different from other
situations in clinical trial design, where only the standard error is
important, not the total sample size by itself, and an important
point to take into account at the trial design stage when using
BIC-type criteria.

In terms of the simulation results we considered two candidate set of
models, one did not include an ANOVA model and one included an ANOVA
model. In the first situation AIC-type criteria overall performed
slightly better than the BIC, which penalized the more complex models too strongly for most
situations. However, when allowing the ANOVA model to be selected as
well, it turned out that the AIC selected it too often in some
situations, leading to a decreased performance. However, over all
scenarios, models and model selection criteria there seemed little
value in adding an ANOVA based model to the set of candidate models,
for the scenarios we evaluated. Approaches that selected this model
more often (like the AIC-type criteria) decreased in performance most,
compared to the candidate set without the ANOVA model.

The most general observation from the simulation is however that the
model averaging methods outperformed the corresponding model selection
methods. Even though the benefit is typically not large, it is
consistent across considered candidate sets, models, designs, total
sample sizes, performance metrics and methods. In terms of which model
averaging method to use (weights based or bootstrap) no clear message
emerges. An advantage of the bootstrap model averaging method from a
pragmatic perspective is that confidence intervals that take into
account model uncertainty, are straightforward to obtain from the
generated bootstrap samples.

\bigskip
\noindent
{\bf Acknowledgements}. This research has received funding from the Collaborative Research Center ``Statistical modeling of nonlinear
dynamic processes'' (SFB 823, Teilprojekt C1, C2) of the German Research Foundation (DFG) and  from the European Union Seventh Framework Programme [FP7 2007-2013]
under grant agreement no 602552 (IDEAL - Integrated Design and Analysis of small population group trials).

\setlength{\bibsep}{1pt}
\begin{small}
\bibliographystyle{apalike}
\bibliography{literature}

\begin{thebibliography}{}

\bibitem[Akaike, 1974]{akai1973}
Akaike, H. (1974).
\newblock A new look at the statistical model identification.
\newblock {\em IEEE Transactions on Automatic Control AC}, 19:716--723.

\bibitem[Bornkamp, 2015]{born2015}
Bornkamp, B. (2015).
\newblock Viewpoint: Model selection uncertainty, pre-specification and model
  averaging.
\newblock {\em Pharmaceutical Statistics}, 14(2):79--81.

\bibitem[Bornkamp et~al., 2007]{phrma:2007}
Bornkamp, B., Bretz, F., Dmitrienko, A., Enas, G., Gaydos, B., Hsu, C.-H.,
  K{\"o}nig, F., Krams, M., Liu, Q., Neuenschwander, B., Parke, T., Pinheiro,
  J.~C., Roy, A., Sax, R., and Shen, F. (2007).
\newblock Innovative approaches for designing and analyzing adaptive
  dose-ranging trials.
\newblock {\em Journal of Biopharmaceutical Statistics}, 17:965--995.

\bibitem[Bornkamp et~al., 2013]{born2013}
Bornkamp, B., Pinheiro, J., and Bretz, F. (2013).
\newblock {\em DoseFinding: Planning and Analyzing Dose Finding experiments}.

\bibitem[Breiman, 1996]{brei1996}
Breiman, L. (1996).
\newblock Bagging predictors.
\newblock {\em Machine Learning}, 24:123--140.

\bibitem[Bretz et~al., 2008]{bret:hsu:pinh:2008}
Bretz, F., Hsu, J.~C., Pinheiro, J.~C., and Liu, Y. (2008).
\newblock Dose finding - a challenge in statistics.
\newblock {\em Biometrical Journal}, 50:480--504.

\bibitem[Bretz et~al., 2005]{bret:pinh:bran:2005}
Bretz, F., Pinheiro, J.~C., and Branson, M. (2005).
\newblock Combining multiple comparisons and modeling techniques in
  dose-response studies.
\newblock {\em Biometrics}, 61:738--748.

\bibitem[Buckland et~al., 1997]{buck1997}
Buckland, S.~T., Burnham, K., and Augustin, N.~H. (1997).
\newblock Model selection: An integral part of inference.
\newblock {\em Biometrics}, 53:603--618.

\bibitem[CHMP, 2014]{ema:2014}
CHMP (2014).
\newblock Qualification opinion of {MCP}-{M}od as an efficient statistical
  methodology for model-based design and analysis of {P}hase {II} dose finding
  studies under model uncertainty.
\newblock European Medicines Agency, Science Medicines Health, Committee for
  Medicinal Products for Human Use (CHMP), EMA/CHMP/SAWP/757052/2013 available
  at http://goo.gl/imT7IT.

\bibitem[Claeskens and Hjort, 2008]{clae2008}
Claeskens, G. and Hjort, N.~L. (2008).
\newblock {\em Model Selection and Model Averaging}.
\newblock Cambridge Series in Statistical and Probabilistic Mathematics.
  Cambridge University Press.

\bibitem[Clyde, 2000]{clyd2000}
Clyde, M. (2000).
\newblock Model uncertainty and health effect studies for partculate matter.
\newblock {\em Environmetrics}, 11:745--763.

\bibitem[Dragalin et~al., 2007]{drag:hsua:padm:2007}
Dragalin, V., Hsuan, F., and Padmanabhan, S.~K. (2007).
\newblock Adaptive designs for dose-finding studies based on the sigmoid emax
  model.
\newblock {\em Journal of Biopharmaceutical Statistics}, 17:1051--1070.

\bibitem[Draper, 1995]{drap1995}
Draper, D. (1995).
\newblock {Assessment and Propagation of Model Uncertainty}.
\newblock {\em Journal of Royal Statisctical Society. Series
  B(Methodological)}, 57(1):45--97.

\bibitem[Hjort and Claeskens, 2003]{hjor2003}
Hjort, N.~L. and Claeskens, G. (2003).
\newblock {Frequentist Model Average Estimators}.
\newblock {\em Journal of the American Statistical Association}, 98:879--899.

\bibitem[Hurvich and Tsai, 1989]{hurv1989}
Hurvich, C. and Tsai, C. (1989).
\newblock {Regression and Time Series Model Selection in Small Samples}.
\newblock {\em Biometrika}, 76:297--307.

\bibitem[Pinheiro et~al., 2014]{pinh:born:glim:2014}
Pinheiro, J.~C., Bornkamp, B., Glimm, E., and Bretz, F. (2014).
\newblock Model-based dose finding under model uncertainty using general
  parametric models.
\newblock {\em Statistics in Medicine}, 33:1646--1661.

\bibitem[Pukelsheim, 2006]{pukelsheim2006}
Pukelsheim, F. (2006).
\newblock {\em Optimal Design of Experiments}.
\newblock SIAM, Philadelphia.

\bibitem[Raftery and Zheng, 2003]{raft:2003}
Raftery, A. and Zheng, Y. (2003).
\newblock Discussion: Performance of bayesian model averaging.
\newblock {\em Journal of the American Statistical Association}, 98:931--938.

\bibitem[Schwarz, 1978]{schw1978}
Schwarz, G. (1978).
\newblock Estimating the dimension of a model.
\newblock {\em The Annals of Statistics}, 6(2):461--464.

\bibitem[Takeuchi, 1976]{take1976}
Takeuchi, K. (1976).
\newblock Distribution of informational statistics and a criterion of model
  fitting.
\newblock {\em Suri-Kagaku(Mathematical Sciences)}, 153:12--18.
\newblock In Japanese.

\bibitem[Thomas, 2006]{thom:2006}
Thomas, N. (2006).
\newblock Hypothesis testing and {B}ayesian estimation using a sigmoid {E}max
  model applied to sparse dose designs.
\newblock {\em Journal of Biopharmaceutical Statistics}, 16:657--677.

\bibitem[Ting, 2006]{ting:2006}
Ting, N. (2006).
\newblock {\em Dose finding in drug development}.
\newblock Spinger.

\bibitem[Verkindre et~al., 2010]{verk2010}
Verkindre, C., Fukuchi, Y., Fl\'{e}male, A., Takeda, A., Overend, T., Prasad,
  N., and Dolker, M. (2010).
\newblock Sustained 24-h efficacy of nva237, a once-daily long-acting
  muscarinic antagonist, in copd patients.
\newblock {\em Respiratory Medicine}, 104:1482--1489.

\bibitem[Verrier et~al., 2014]{verr:2014}
Verrier, D., Sivapregassam, S., and Solente, A.-C. (2014).
\newblock Dose-finding studies, mcp-mod, model selection, and model averaging:
  Two applications in the real world.
\newblock {\em Clinical Trials}, 11:476--484.

\bibitem[Wassermann, 2000]{wass2000}
Wassermann, L. (2000).
\newblock Bayesian model selection and model averaging.
\newblock {\em Journal of Mathematical Psychology}, 44:92--107.

\bibitem[White, 1982]{white82}
White, H. (1982).
\newblock {Maximum Likelihood Estimation of Misspecified Models}.
\newblock {\em Econometrica}, 50:1--26.

\end{thebibliography}
\end{small}

\appendix
\section{Appendix: Background on model selection criteria based on the AIC}
\label{sec:append-backgr-aic}

As in Section~\ref{sec:not}, let $\hat{\theta}_\ell$ denote the ML estimator in model $\cM_\ell$ ($\ell=1, \ldots, L$). As shown by \cite{white82} this estimator converges  in probability  to the KL-divergence minimizing parameter  $\theta^*_\ell$ under certain regularity conditions. \\
This gives the estimated KL-divergences
$$KL(p_{\ell}(\cdot | \cdot, \hat{\theta}_\ell), g)= \sum_{i=1}^{k} \frac{n_i}{N} \int \log\left( \frac{g(y|d_i)}{p_{{\ell}}(y| d_i, \hat{\theta}_{\ell})} \right) g(y|d_i) d y, \ell= 1, \ldots, L.$$
Note that this term is a random variable with expected value
$$\sum_{i=1}^{k} \frac{n_i}{N} \left( \int  g(y|d_i) \log (g(y|d_i)) dy - \E \left[ \int g(y|d_i) \log p_{\ell}(y| d_i, \hat{\theta}_{\ell} )dy \right] \right)$$
because here the estimator $\hat{\theta}_{\ell}$ is considered as fixed.
Both the first and the second term within the sum depend on the true density $g(y|d)$, whereas only the second one depends on the considered model $\cM_{\ell}$ and its estimator $\hat{\theta}_{\ell}$. Thus, we only need to estimate the term
$$ Q_N(\cM_{\ell}):= \E [R_n]:= \E \left[ \sum_{i=1}^k \frac{n_i}{N} \int g(y|d_i) \log p_{{\ell}}(y| d_i, \hat{\theta}_{\ell})dy \right] $$
in order to distinguish the quality of approximations. \\ 

For the estimation of $Q_N(\cM_{\ell})$ we replace the expected value and integral by the mean depending on the observations: Thus, an estimator for $Q(\cM_{\ell})$ is given by
\begin{equation*}\label{eq:llf}
\hat{Q}_N(\cM_\ell)= \frac{1}{N} \sum_{i=1}^{k} \sum_{j=1}^{n_i} \log p_{\ell}(Y_{ij}|d_i, \hat{\theta}_{\ell})= \frac{1}{N} \max_{\theta_\ell} \log(\mathcal{L}_N(\cM_\ell, \theta_\ell)),
\end{equation*}
where $\mathcal{L}_N(\cM_{\ell}, \theta_x{\ell})$ is the likelihood function of model $\cM_{\ell}$ evaluated at the parameter $\theta_{\ell}$. In principle a model could be chosen from $\cM_1, \ldots, \cM_L$ which leads to the largest value of $\hat{Q}_N(\cM_{\ell})$ (${\ell}=1, \ldots, 5$). However, this naive estimator usually chooses the model with the largest number of parameters which often leads to an overfit of the data. This property is a consequence of the fact that the log likelihood function is an increasing function of the dimension $d_{\cM}$ of the parameter $\theta_{\cM}$. It is even possible to calculate the approximate bias (see \cite{clae2008}) as
\begin{equation*}
\E[\hat{Q}_N(\cM_{\ell})] - Q_N(\cM_{\ell}) \approx \frac{pen^\star_{{\ell}}}{N},
\end{equation*}
where $pen^\star_{{\ell}}= tr(K(\cM_{\ell})J^{-1}(\cM_{\ell}))$,
$$K(\cM_{\ell})=\sum_{i=1}^{k} \frac{n_i}{N} \E \left[\frac{\partial \log p_{{\ell}}(Y|d_i, \theta_{\ell}) }{\partial \theta_{\ell}} \left( \frac{\partial \log p_{{\ell}}(Y|d_i, \theta_{\ell})}{\partial \theta_{\ell}}\right)^T\right]$$
denotes the Fisher information matrix and the matrix $J$,
$$J^{-1}(\cM_{\ell}) = -\left( \sum_{i=1}^k \frac{n_i}{N} \E \left[ \frac{\partial^2 \log p_{{\ell}}(Y|d_i, \theta_{\ell})}{\partial^2\theta_{\ell}}\right]\right)^{-1},$$
the negative inverse of the expectation of the second derivative of the log likelihood function.
If the considered density $p_{{\ell}}$ and the true density $g$ coincide (i.e. model $p_{\ell}$ is the true one) and certain regularity conditions (c.f. \cite{white82}) are fulfilled, we have $K(\cM_{\ell})=J(\cM_{\ell})$ and consequently $pen^*_{{\ell}}= d_{\cM_{\ell}}$, where $d_{\cM_{\ell}}$ denotes the dimension of the parameter $\theta_{{\ell}}$.

In conclusion, a bias corrected estimator for the second part of the KL-divergence is given by \eqref{eq:est}.
As outlined in Section~\ref{sec:aic}, the AIC-based criteria from Table \ref{tab:msc}
are based on this estimator $Q^*_N(\cM_{\ell})$ using different estimators for the penalty term $pen_{{\ell},I}= pen^\star_{{\ell}}$.
Note that for the TIC the two matrices $K$ and $J^{-1}$ and thus $p^\star_{\cM_{\ell}}$ are explicitly estimated by
$$\hat{K}(\cM_{\ell})=\sum_{i=1}^k \sum_{j=1}^{n_i} \frac{1}{n_i}\frac{\partial \log p_{\ell}(y_{ij}|d_i, \hat{\theta}_{\ell}) }{\partial \theta_{\ell}} \left( \frac{\partial \log p_{{\ell}}(y_{ij} |d_i,\hat{\theta}_{\ell})}{\partial \theta_{\ell}}\right)^T,$$
\mbox{ and}
$$\hat{J}(\cM_{\ell}) = -\sum_{i=1}^k \sum_{j=1}^{n_i} \frac{1}{n_i} \frac{\partial^2 \log p_{{\ell}}(y_{ij}|d_i, \hat{\theta}_{\ell})}{\partial^2\theta_{\ell}},$$
respectively. The resulting penalty term is therefore given by $\trace(\hat{J}^{-1}(\cM_{\ell}) \hat{K}(\cM_{\ell}) )$ \cite{take1976}.

\newpage

\section{Selection Probabilities}\label{sec:prob}
In this Section the probabilities that a selection criterion chooses a response model given a specific scenario are displayed in the case where the candidate models are given by the linear, the quadratic, the Emax and the Sigmoid Emax model.

\begin{figure}[h!]
   \subfigure{\includegraphics[width=0.45 \textwidth]{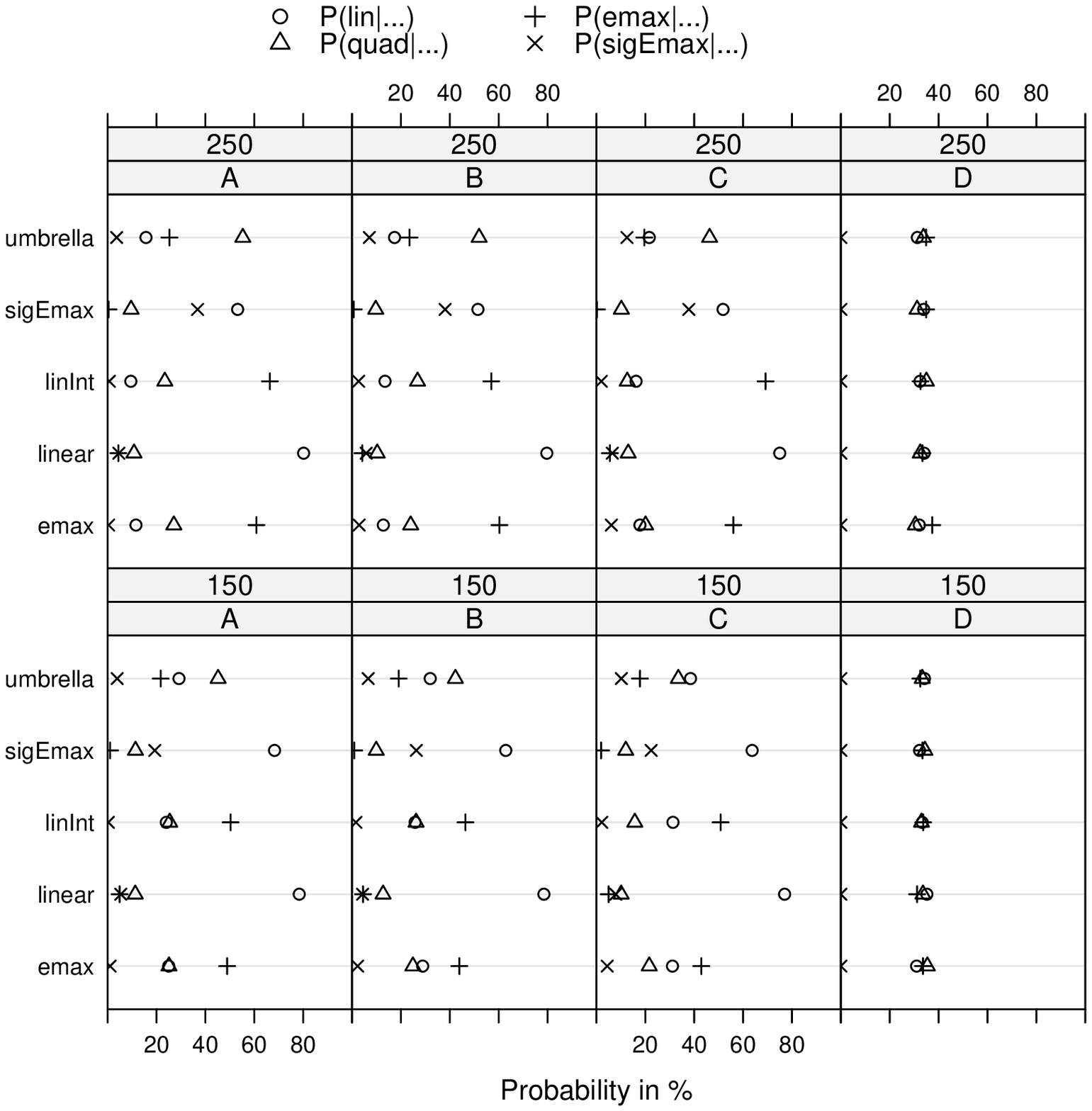}}\,
   \subfigure{\includegraphics[width=0.45 \textwidth]{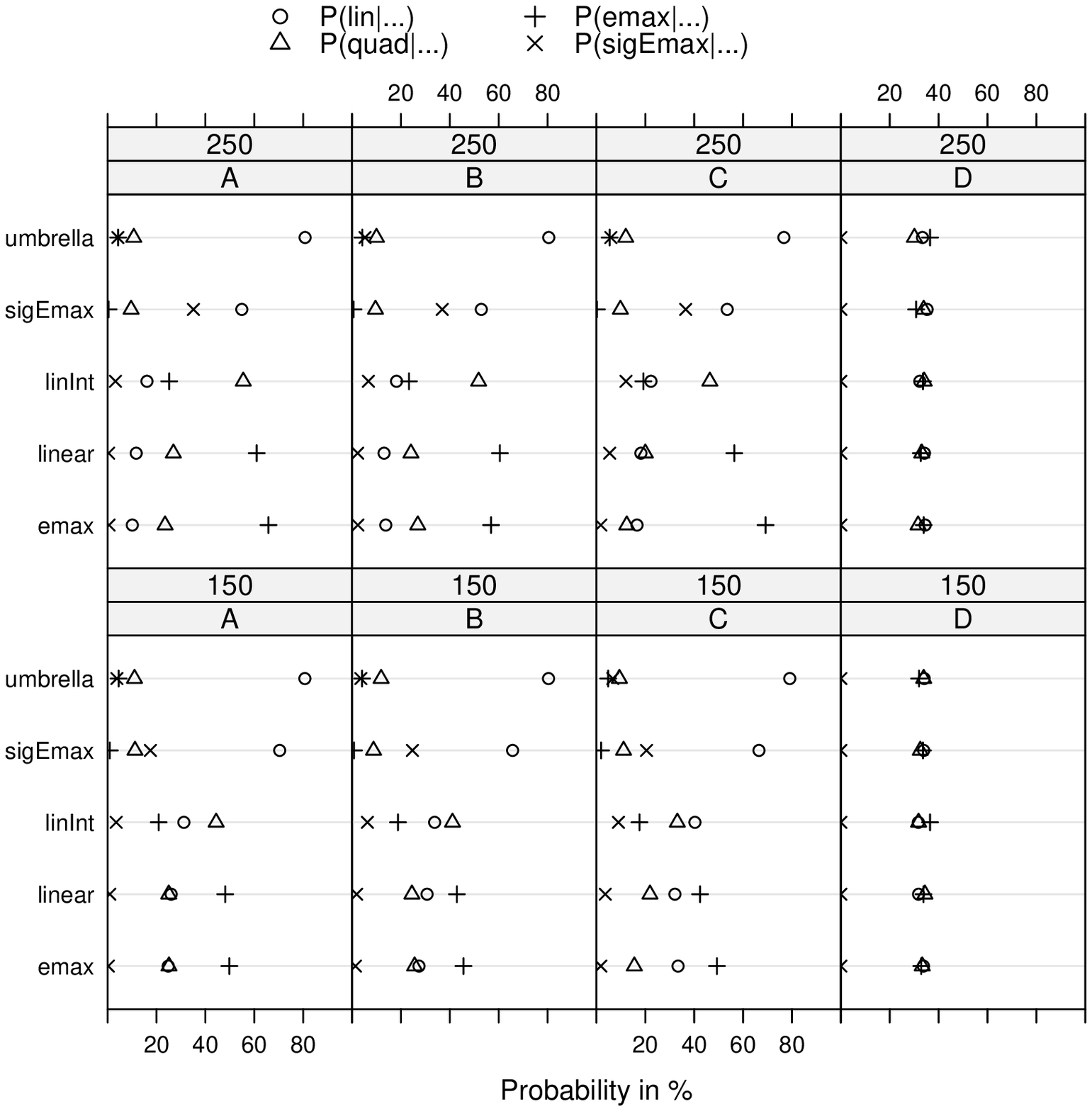}}
    \caption[Selection Probabilities AIC, $\AIC_C$]{\emph{The probability that the $AIC$ (left) and the $AIC_C$ (right) choose a response model given a scenario.}}
    \label{fig:auswahl_aic_aicc}
\end{figure}
\begin{figure}[h!]
   \subfigure{\includegraphics[width=0.45 \textwidth]{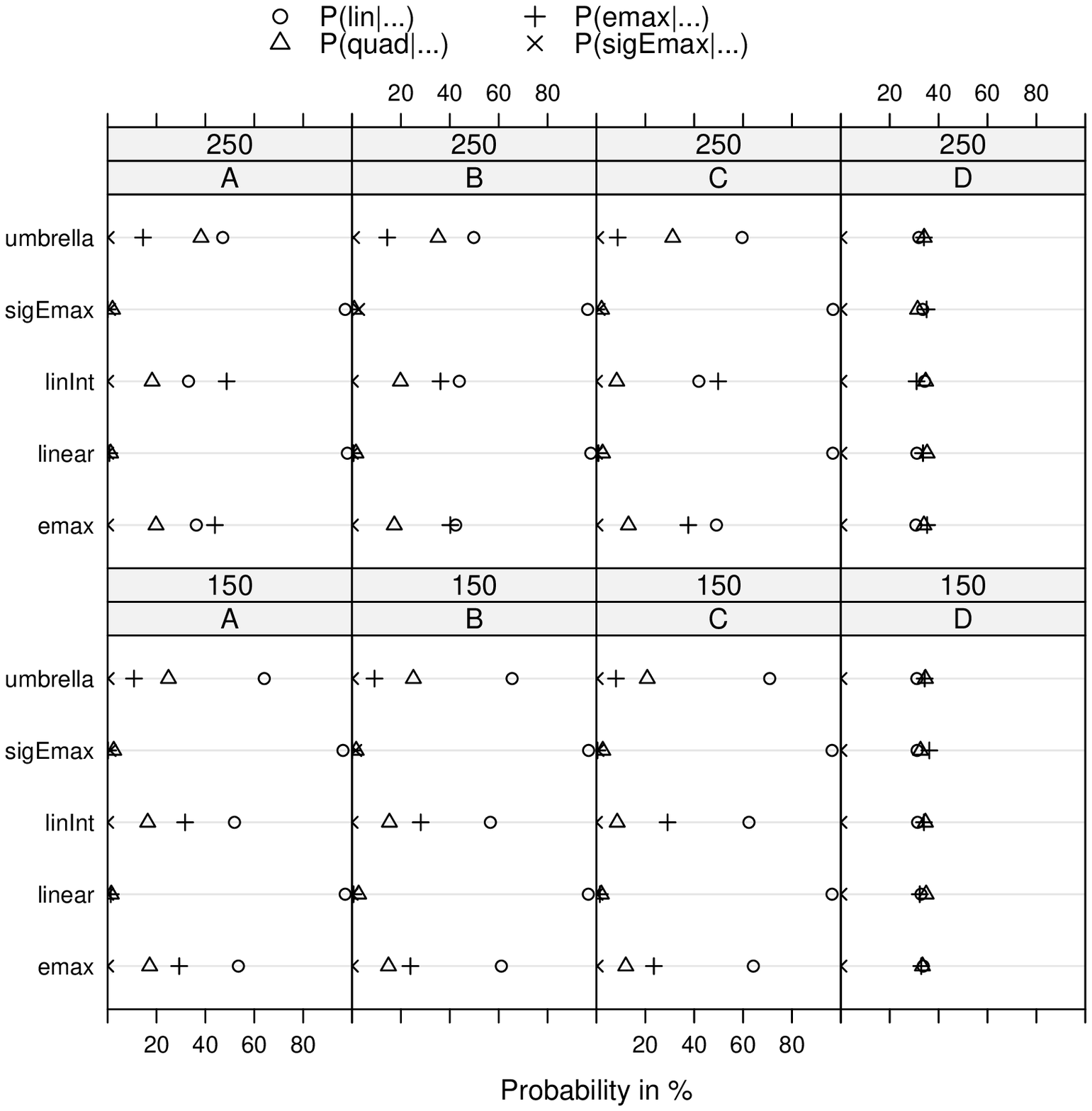}}\,
   \subfigure{\includegraphics[width=0.45  \textwidth]{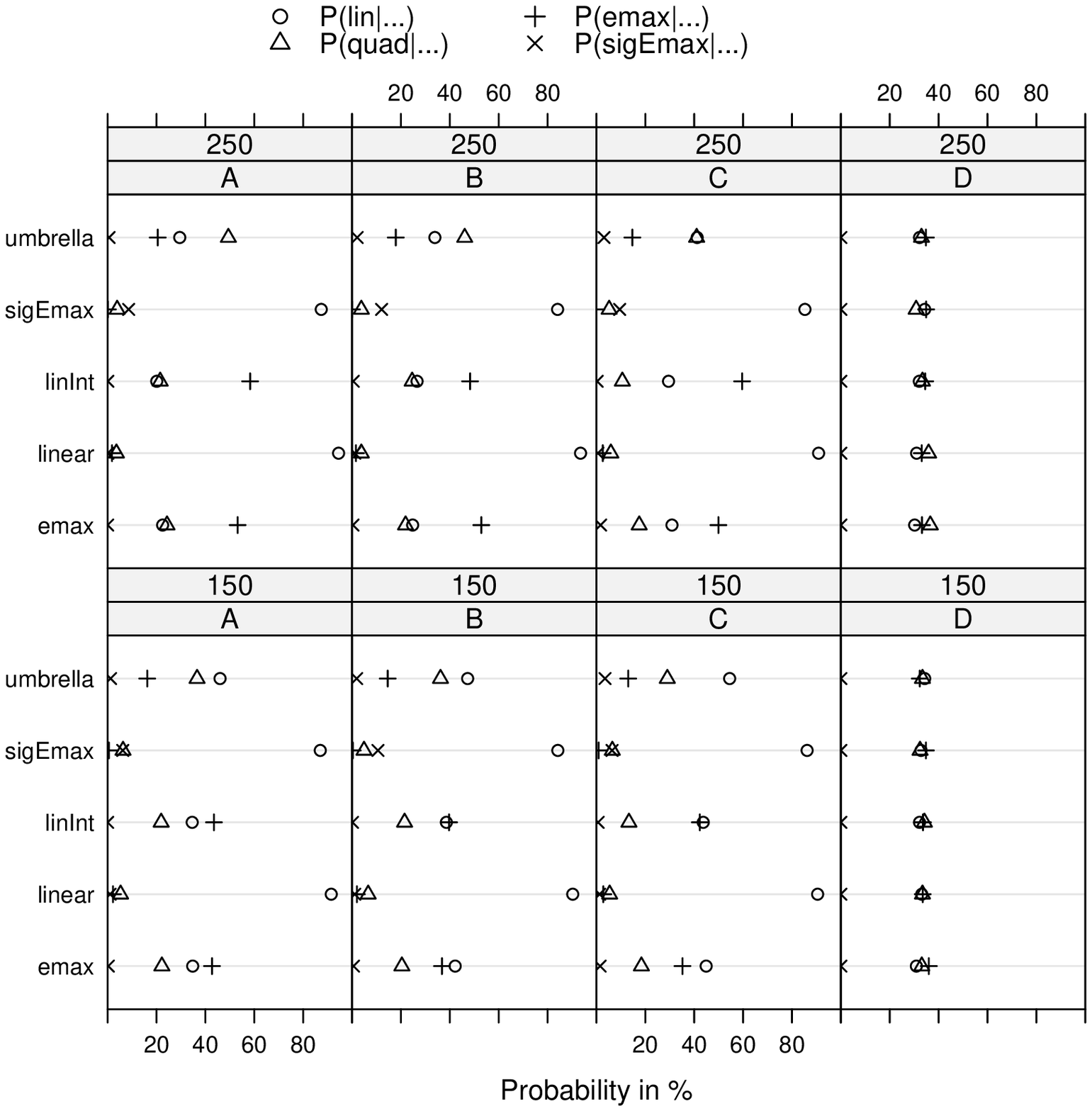}}
    \caption[Selection Probabilities BIC, $\BIC_2$]{\emph{The probability that the $BIC$ (left) and the $BIC_2$ (right) choose a response model given a scenario.}}
    \label{fig:auswahl_bic_bic2}
\end{figure}
\begin{figure}[h!]
\centering
\includegraphics[width=0.48 \textwidth]{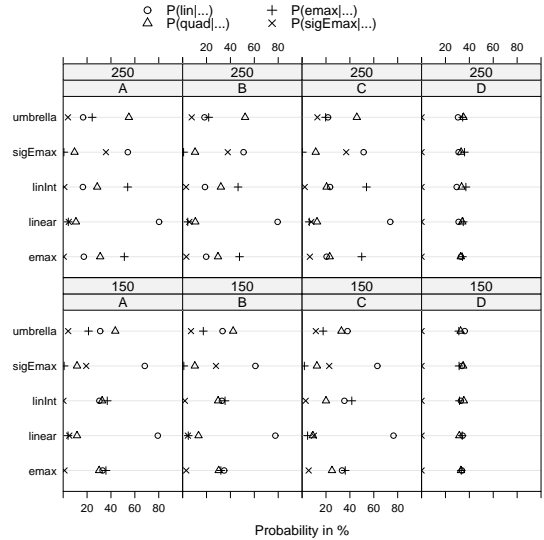}
\caption[Selection Probabilities TIC]{\emph{The probability that the $TIC$ chooses a response model given a scenario.}}
\label{fig:auswahl_tic}
\end{figure}

\newpage

\section{Tables for Simulation Study with ANOVA}\label{sec:withanova}
\begin{table}[h!]
\centering
\begin{tabular}{|l | ccccc|}
\hline
    &       AIC  &     BIC  &    $\BIC_2$ &       TIC &    $\AIC_C$ \\
\hline
    Probability & 41\% & 34\% & 39\% & 41\% & 22\% \\
ASMSE(A) & 1.52 & 1.55 & 1.45 & 1.52 & 1.57 \\
ASMSE(B) & 1.57 & 1.57 & 1.47 & 1.56 & 1.60  \\
ASMSE(C) &1.50 & 1.47 & 1.40 & 1.50 & 1.53 \\
ASMSE(D) &1.51 & 1.59 & 1.47 & 1.51 & 1.59 \\
$\mbox{ASMSE}_{\mbox{td}} $&1.73 & 2.43 & 1.96 &1.74  & 2.74 \\
\hline
\end{tabular}
\caption[Averages MAV]{\textit{The averages of the standardized mean squared errors taken over all scenarios for model selection. ANOVA is among the candidate models.}}
\label{tab:averages_ms_wA}
\end{table}

\begin{table}[h!]
\centering
\begin{tabular}{|l|ccccccc|}
\hline
       &    AIC   &   BIC    & $\BIC_2$    &  TIC  &   $\AIC_C$ & AIC-Boot & BIC-Boot \\
\hline
ASMSE(A)& 1.36 & 1.39 & 1.31 & 1.36 & 1.29 & 1.56 & 1.30 \\
ASMSE(B) & 1.39 & 1.40 & 1.32 & 1.39 & 1.30 & 1.63 & 1.31 \\
ASMSE(C) & 1.34 & 1.33 & 1.27 & 1.34 & 1.28 & 1.54 & 1.25 \\
ASMSE(D) & 1.36 & 1.42 & 1.32 & 1.36 & 1.30 & 1.53 & 1.32\\
$\mbox{ASMSE}_{\mbox{td}}$ & 1.55 & 1.94 & 1.64 & 1.55 & 1.87 & 1.28 & 1.43 \\
\hline
\end{tabular}
\caption[Averages MAV]{\textit{The averages of the standardized mean squared errors taken over all scenarios for model averaging and bootstrap model averaging. ANOVA is among the candidate models.}}
\label{tab:averages_mav_wA}
\end{table}


\begin{figure}[h!]
\begin{center}
\includegraphics[width=0.8\textwidth]{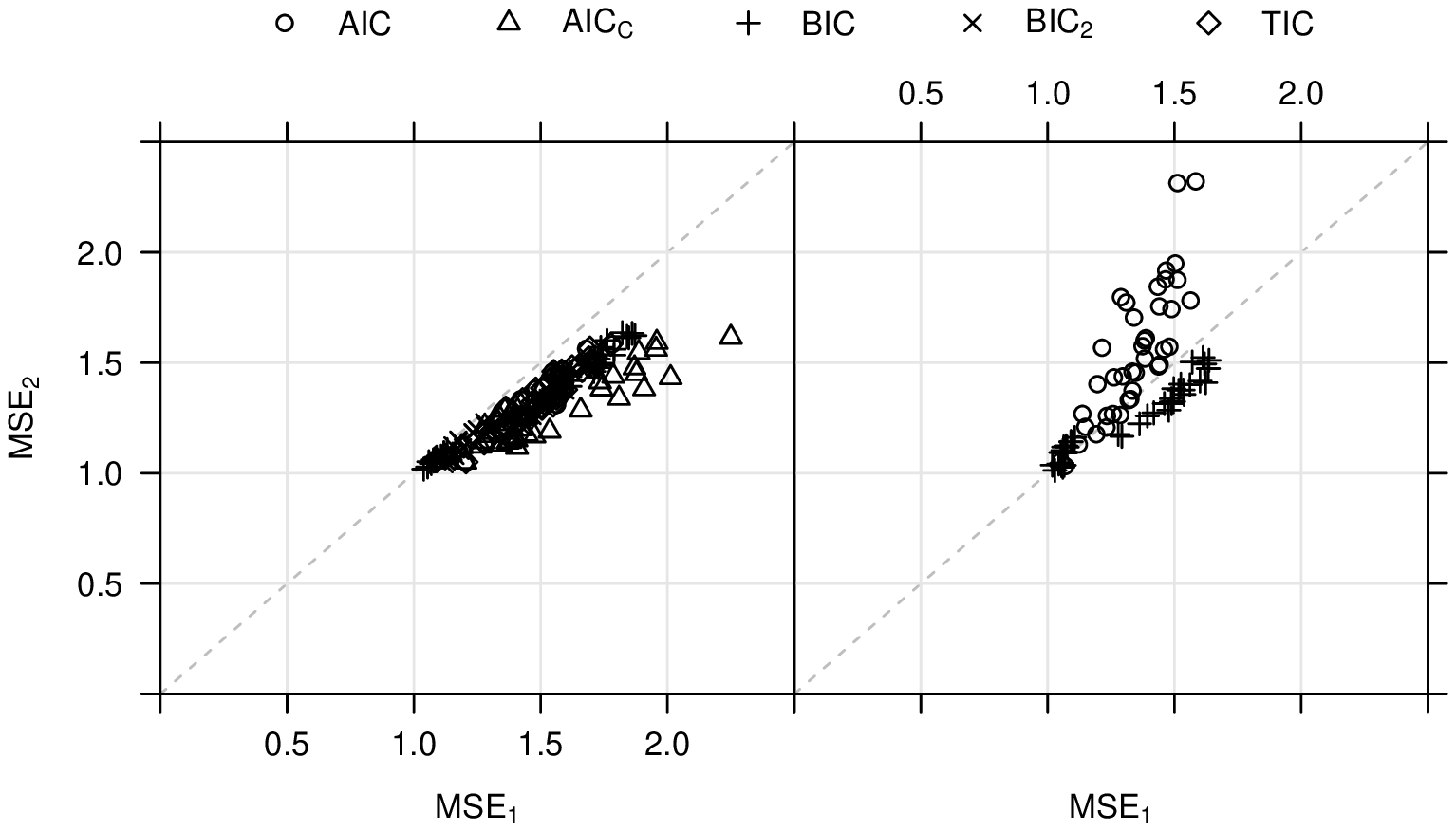}
\end{center}
    \caption[Comparison Dose Response]{\emph{Comparison of model
        selection, model averaging and bootstrap for estimating the
        dose effects in design $C$. The Figure shows the $\mbox{SMSE}$
        values. Left panel: model selection versus model
        averaging. Right panel: model averaging versus bootstrap model
        averaging. ANOVA is among the candidate models.}}
    \label{fig:comp_de_mse_wA}
\end{figure}

\begin{figure}[H]
\begin{center}
\includegraphics[width=0.8\textwidth]{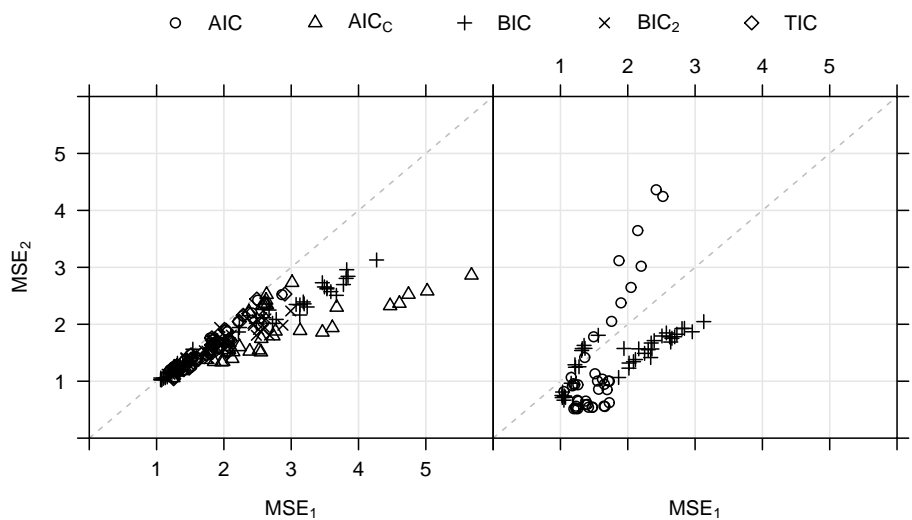}
\end{center}
    \caption[Comparison Target Dose]{\emph{Comparison of model
       selection, model averaging and bootstrap model averaging for estimating the
       target dose. The Figure shows the $\mbox{SMSE}$. Left panel:
       model selection versus model averaging. Right panel: weights
       based model averaging versus bootstrap model averaging. ANOVA is among the candidate models.}}
    \label{fig:comp_td_mse_wA}
\end{figure}

\newpage

\section{Additional Plots for the Example} \label{sec:modfitex}
\begin{figure}[h!]
\subfigure{\includegraphics[width=0.45 \textwidth]{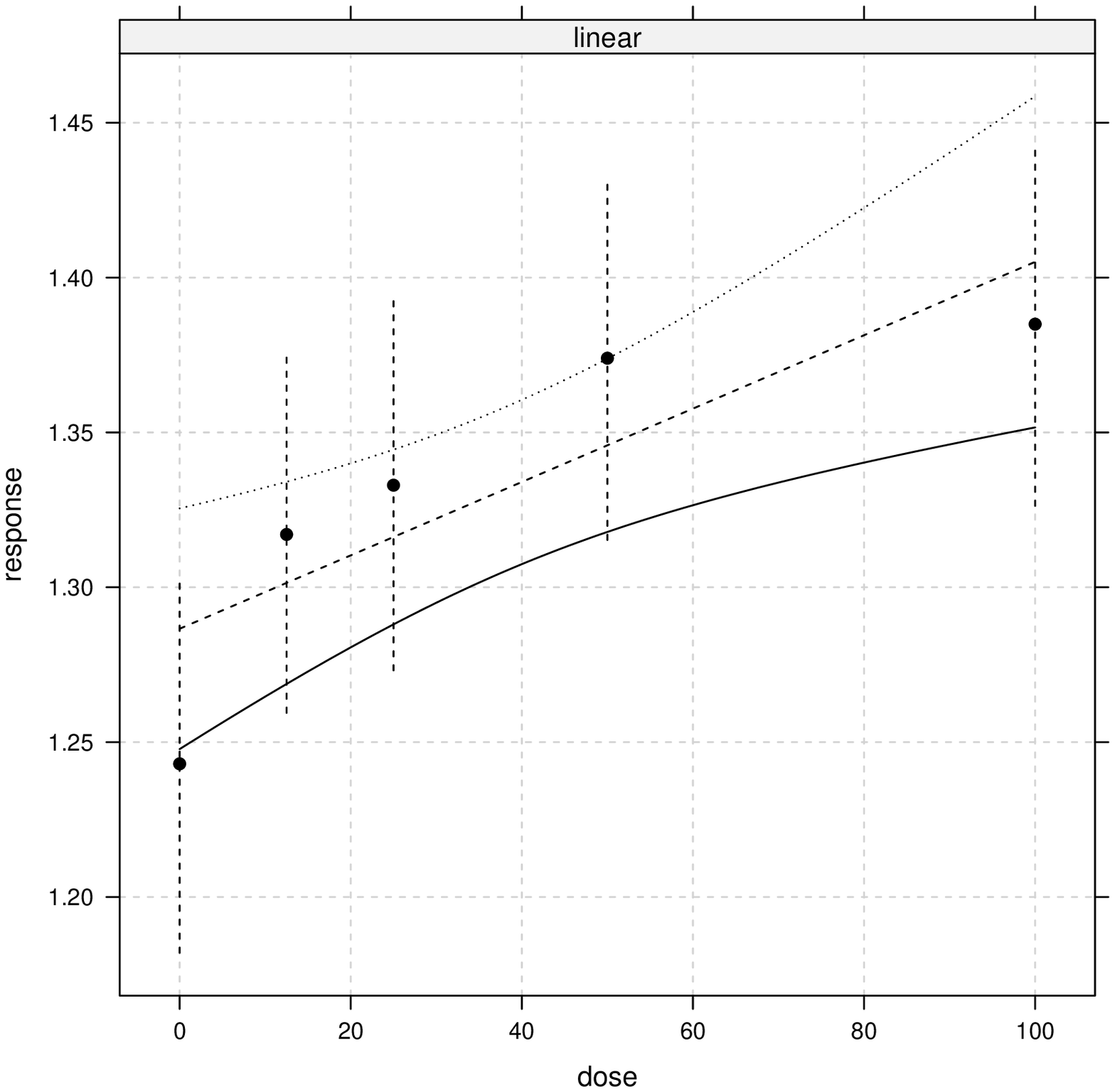}} \, \subfigure{\includegraphics[width=0.45 \textwidth]{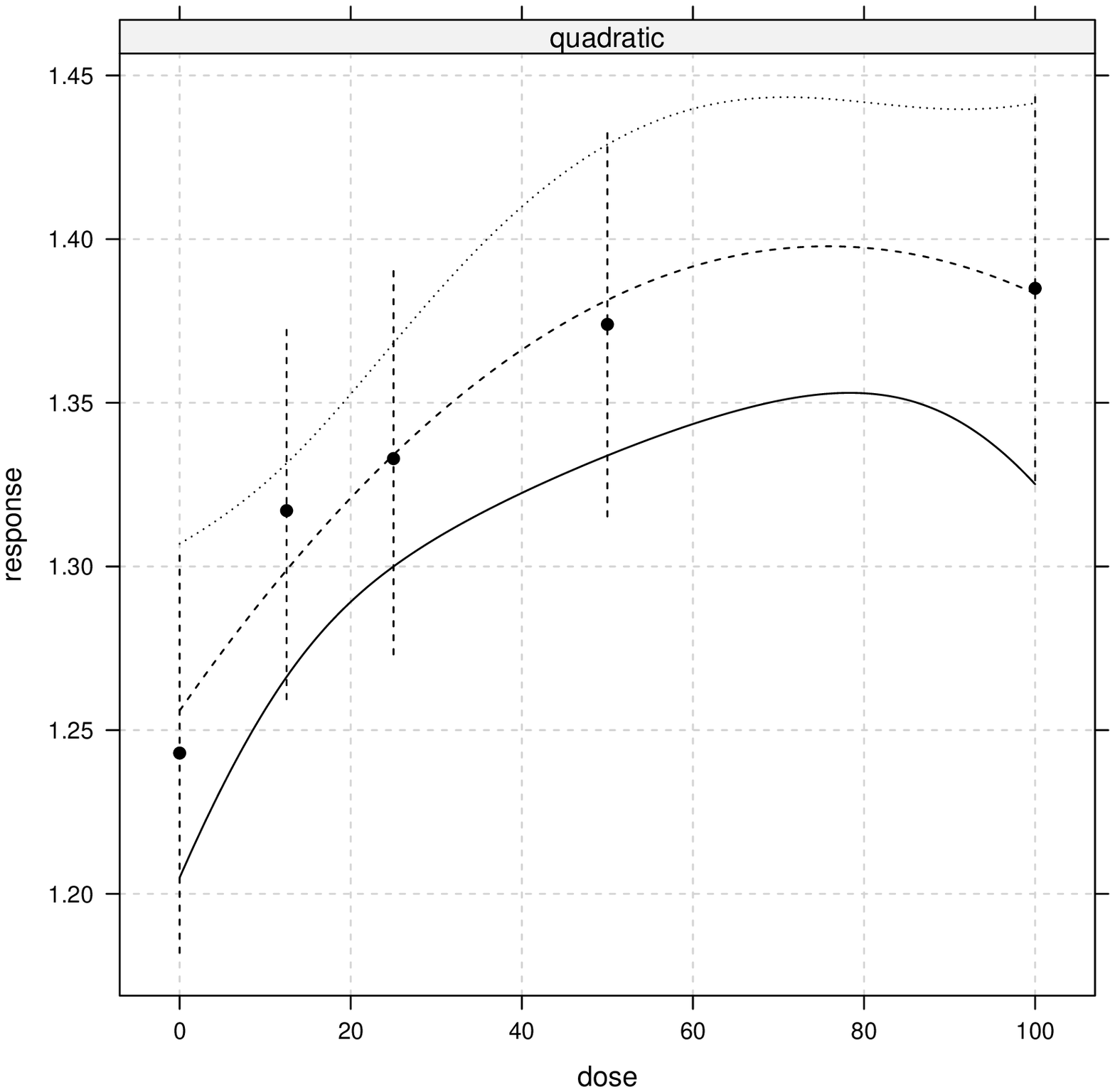}} \\
\subfigure{\includegraphics[width=0.45 \textwidth]{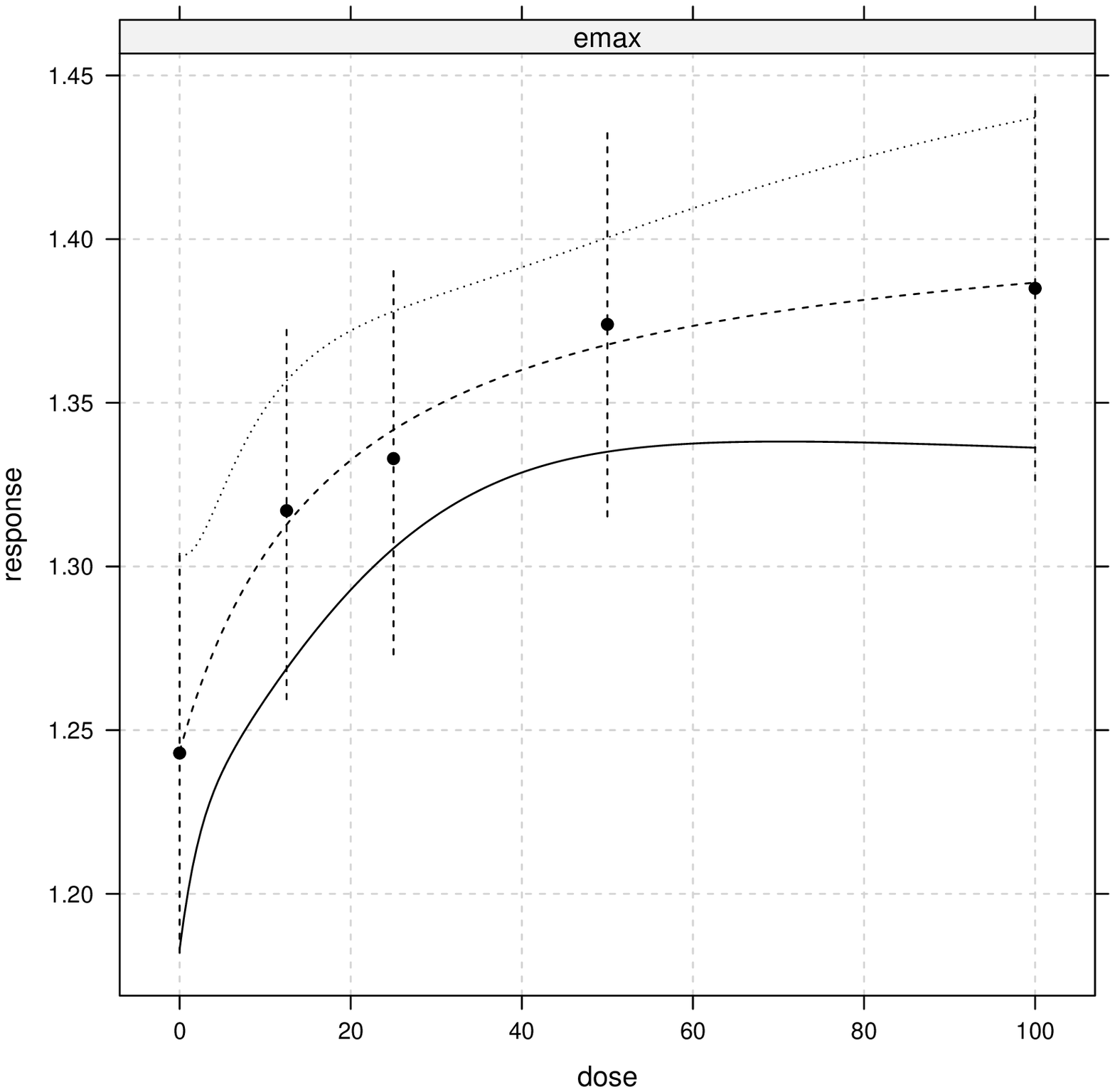}} \, \subfigure{\includegraphics[width=0.45 \textwidth]{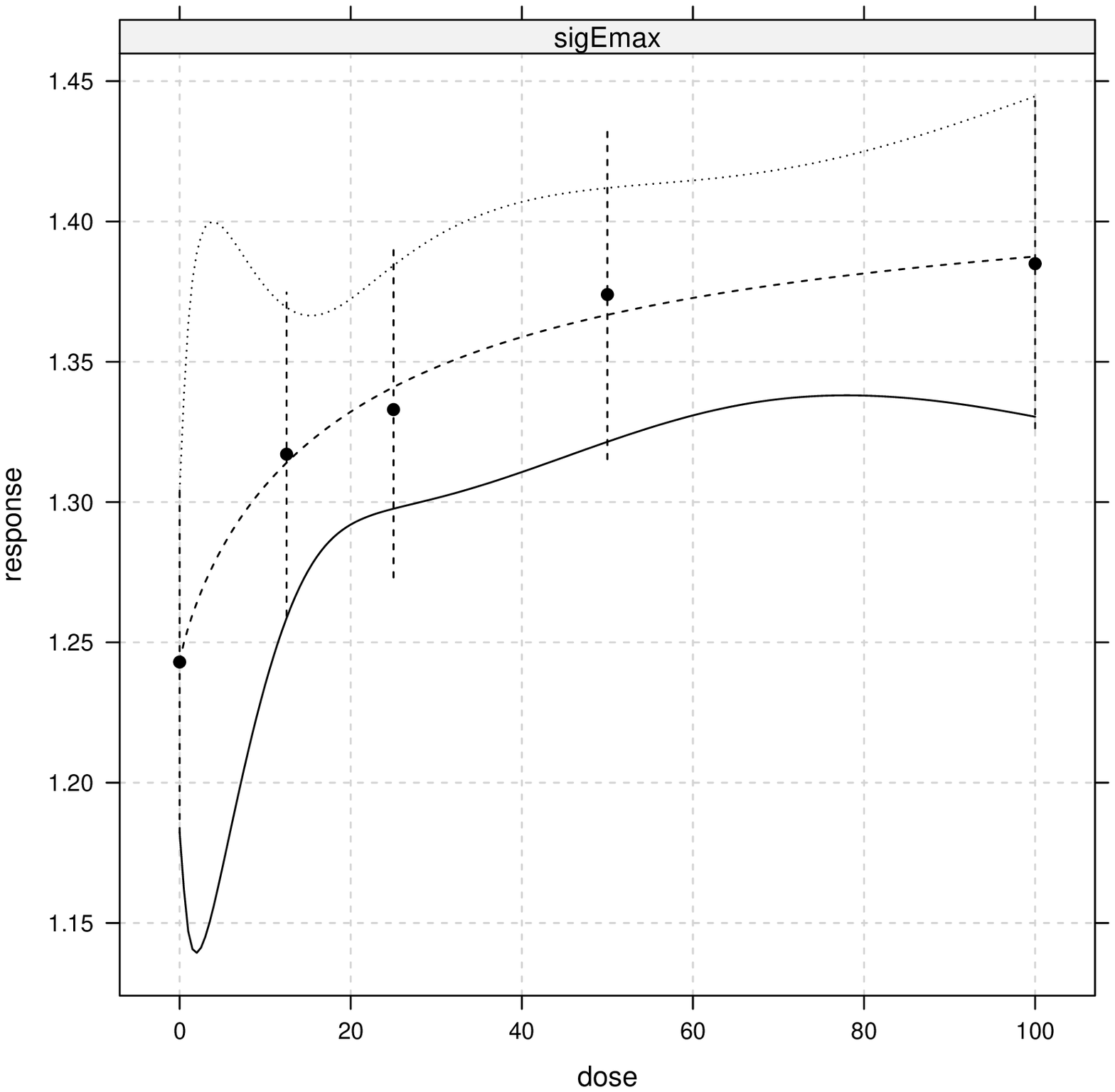}} \\
\caption[Fitted Models] {\emph{The fitted candidate models, namely the linear, the quadratic, the Emax, the Sigmoid Emax and the ANOVA model. The ANOVA model is given by the means at the different dose levels (which are given by the points in every figure).}}
\label{fig:fitmod}
\end{figure}

\end{document}